\documentclass[twocolumn,secnumarabic,amssymb, nobibnotes]{revtex4-2}

\usepackage{lineno}

\usepackage{amsmath,amssymb}
\usepackage{type1cm}
\usepackage{mathrsfs}
\usepackage{url}
\usepackage{graphicx}
\usepackage{color}
\usepackage{booktabs}
\allowdisplaybreaks[1]
\newcommand\HL[1]{{\color{black}#1}}
\newcommand\HLLLLL[1]{{\color{black}#1}}

\usepackage{bm}
\newcommand{\vect}[1]{\boldsymbol{\mathbf{#1}}}
\def\vec#1{\vect{#1}}

\begin{document}
\title{Particle approximation of the two-fluid model for superfluid $\rm ^{4}He$ using smoothed particle hydrodynamics}

\author{Satori Tsuzuki}
\email[The University of Tokyo:]{tsuzuki@jamology.rcast.u-tokyo.ac.jp\\ \url{https://www.satoritsuzuki.org/} }
\affiliation{Research Center for Advanced Science and Technology, The University of Tokyo, 4-6-1, Komaba, Meguro-ku, Tokyo 153-8904, Japan}
\begin{abstract}
This paper presents a finite particle approximation of the two-fluid model for liquid $\rm ^{4}He$ using smoothed particle hydrodynamics (SPH). \HLLLLL{In recent years, several studies have combined the vortex filament model (VFM), which describes quantized vortices in superfluid components, with the Navier--Stokes equations, which describe the motion of normal fluids. These studies led us to assume that coupling both components of the two-fluid model instead of using the VFM to describe the superfluid component enables us to approximate the system. In this study, we formulated a new SPH model that simultaneously solves both equations of motion of the two-fluid model. We then performed a numerical simulation of the rotating liquid $\rm ^{4}He$ using our SPH. The results showed that} the two major phenomena, the emergence of multiple independent vortices parallel to the circular axis and that of the so-called rigid-body rotation, can be reproduced by solving the two-fluid model using SPH. This finding is interesting because it was previously assumed that only a single vortex emerges when addressing similar problems without considering quantum mechanics. \HLLLLL{Our further analysis found that the emergence of multiple independent vortices can be realized by reformulating the viscosity term of the two-fluid model to conserve the angular momentum of the particles around their axes.} Consequently, our model succeeded in reproducing the phenomena observed in quantum cases, even though we solve the phenomenological governing equations of liquid $\rm ^{4}He$.
\end{abstract}
\maketitle

%\linenumbers
\section{Introduction} \label{sec:intro}
The bizarre behavior of liquid helium 4 (hereinafter, $\rm ^{4}He$) has attracted the attention of condensed matter physicists for many years; however, the detailed dynamics of this fluid remain a mystery even now, eighty years after its discovery.

\begin{figure}[t]
\vspace{-1.5cm}
%\begin{center}
\includegraphics[width=0.65\textwidth, clip, bb= 0 0 960 720]{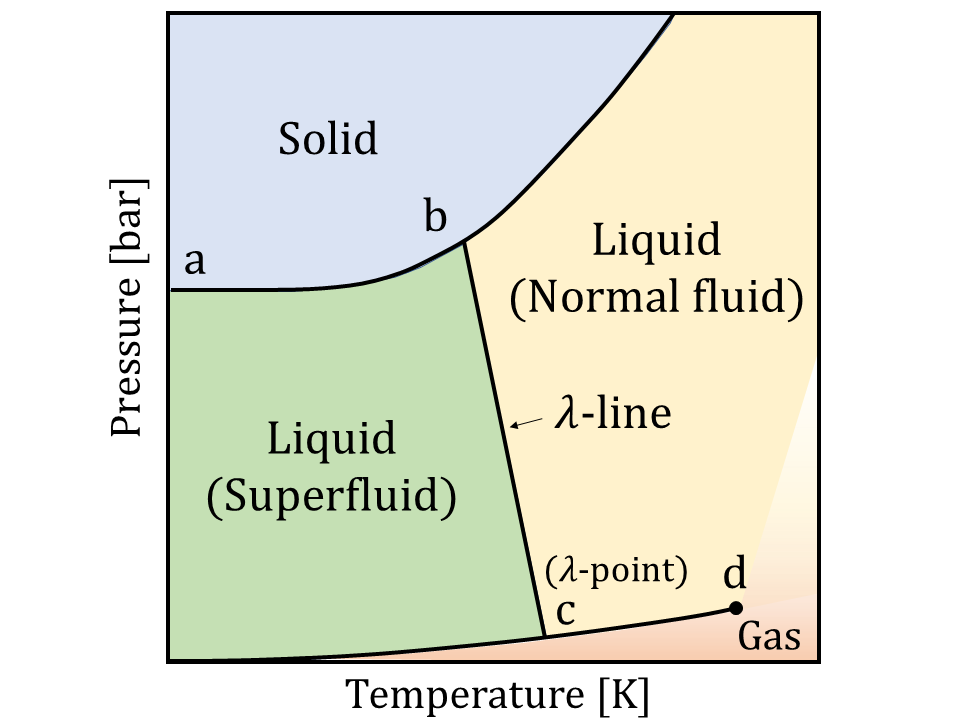}
% \centerline{\includegraphics[scale=0.3]{Figure_SchemDiagram.eps}}
%\end{center}
\caption{Schematic phase diagram of $\rm ^{4}He$. Points $a$, $b$, $c$, and $d$ correspond to the following (p, T) values, respectively: $\rm (25 bar, 0 K)$, $\rm (30 bar, 1.77 K)$, $\rm (50.5 mbar, 2.17 K)$, and $\rm (2.3 bar, 5.22 K)$.}
\label{fig:SchemDiagram}
\end{figure}

Figure~\ref{fig:SchemDiagram} \HLLLLL{presents an overview of} the basic properties of $\rm ^{4}He$ \HLLLLL{in the form of} a schematic phase diagram. 
In 1937, P.~L.~Kapitza found that the viscosity of liquid $\rm ^{4}He$ \HLLLLL{decreased} upon cooling and almost \HLLLLL{reached} zero below the $\lambda$-point (\HLLLLL{labeled} point $c$ in Fig.~\ref{fig:SchemDiagram}). 
Liquid $\rm ^{4}He$ was later confirmed to display viscosity in the high-temperature area \HLLLLL{above} the $\lambda$-line, and to lose it in the low-temperature area below the $\lambda$-line. Once lack of viscosity is achieved in this latter temperature range, liquid $\rm ^{4}He$ displays many unexpected properties that characterize it as a ``superfluid.''
\HLLLLL{By} contrast, liquid $\rm ^{4}He$ within the high-temperature area \HLLLLL{above} the $\lambda$-line is still classified as a ``normal fluid.'' 

The superfluidity condition is characterized by properties such as \HLLLLL{those in} the following examples. Superfluid $\rm ^{4}He$ can penetrate a high-density capillary filter (e.g., gypsum), \HLLLLL{owing} to the so-called superleak effect, whereas this is not possible for normal fluid $\rm ^{4}He$. 
\HLLLLL{Liquid $\rm ^{4}He$ in a vessel equipped with a capillary filter at its bottom and an opening at its top, and dipped in a reservoir, gushes from the top of the vessel because the surrounding superfluid $\rm ^{4}He$ flows back into the container owing to the superleak effect. This process is known as the ``fountain effect.''} In addition, the ``film flow effect'' takes place when superfluid $\rm ^{4}He$ spontaneously comes out of its vessel as a result of \HLLLLL{the} van der Waals forces between \HLLLLL{the} helium atoms and walls being stronger than the forces between \HLLLLL{the} $\rm ^{4}He$ atoms themselves.

Many physicists have made efforts to explain these interesting properties of superfluids at both the ontological and phenomenological level\HLLLLL{s}. One year after the discovery by P.~L.~Kapitza, F.W.~London explained the phase transition of liquid $\rm ^{4}He$ from a quantum mechanical \HLLLLL{standpoint}, comparing it with a type of Bose--Einstein condensation~(BEC)~\cite{Bose1924}. London's theory has been validated by many experiments.
\HLLLLL{Apart from this}, the ``two-fluid model'' proposed by L.~Tisza~\cite{TISZA1938} and L.~Landau~\cite{PhysRev.60.356}, which assumes the coexistence of superfluidity and normal fluidity in liquid $\rm ^{4}He$, has become the most acknowledged phenomenological description.
A remarkable achievement of the two-fluid model is that it connects the macroscopic perspective of thermodynamics with quantum mechanics, \HLLLLL{such} that the superfluid can be described within \HLLLLL{the} framework of basic mechanics~(\HLLLLL{for detailed descriptions of the two-fluid model, see Section~\ref{seq:twofluidmodel}}). 
%To be more specific, 

From the viewpoint of numerical \HLLLLL{analysis,} a finite approximation of \HLLLLL{the phenomenological} governing equations represents an essential step toward \HLLLLL{the} direct numerical simulations \HLLLLL{of large-scale problems on a continuum mechanical scale, \HLLLLL{for example}, the fountain effect or film-flow effect.} However, to date, only a few studies have adopted such an approach; this is primarily because most previous \HLLLLL{studies took} a \HLLLLL{quantum mechanical standpoint}~(e.g., numerical \HLLLLL{analyses} of nonlinear $\rm Schr\ddot{o}dinger$ equations~\cite{LEHTOVAARA200478, DANAILA20106946}, and quantum Monte Carlo calculations~\cite{CEPERLEY555, PhysRevB.52.3654} \HLLLLL{with the aim of examining} the static responses~\cite{PhysRevLett.69.1837} or to explore quantum vortices and turbulence~\cite{PhysRevB.54.1205, RevModPhys.67.37, donnelly1991quantized, PhysRevLett.95.145302, PhysRevB.89.224516}). \HLLLLL{Conversely,} previous \HLLLLL{studies}~\cite{doi:10.2514/3.149, MURAKAMI1991443, Bottura_2009, darve2011phenomenological} introduced direct numerical simulations of the two-fluid model. \HLLLLL{Nevertheless,} they focused on \HLLLLL{mesh- or grid-based approaches}; some studies~\cite{doi:10.2514/3.149, MURAKAMI1991443} \HLLLLL{involved the development of} finite difference approximations, and others~\cite{Bottura_2009, darve2011phenomenological} \HLLLLL{led to} finite element approximations. 
\HLLLLL{Notably}, none of the previous studies reported a finite particle approximation of \HLLLLL{the two-fluid model}.

Numerical simulations using the finite particle approximation are \HLLLLL{advantageous in many respects} because of the discretization based on many-particle interacting systems. Each particle moves in space \HLLLLL{while} interacting with every other particle, taking \HLLLLL{on} either the state of a superfluid or \HLLLLL{that of a} normal fluid. The mechanical picture, in this case, is much closer to the view of quantum statistical systems that follow Bose--Einstein condensates, compared \HLLLLL{with} the case of grid-based methods. Using the same number of particles as real atoms in liquid $\rm ^{4} He$ in simulations remains challenging even today. Nevertheless, finite particle approximation has a high possibility of bridging the gulf between the phenomenological and microscopic concepts of liquid $\rm ^{4} He$. \HLLLLL{For example,} it \HLLLLL{should} be possible to devise numerical models that replicate quantum effects such as vortex reconnections and incorporate them into the discretized two-fluid model. Furthermore, from a practical \HLLLLL{point of view}, the particle approximation is capable of handling complex boundaries. It has great potential to perform simulations \HLLLLL{that are sufficiently} realistic to reproduce a \HLLLLL{large-scale} experimental setup. However, in addition to these potential advantages, the computational costs are much higher than \HLLLLL{those of} other methods. \HLLLLL{Furthermore}, because the non-uniformity of particle distributions often causes \HLLLLL{the} numerical accuracy \HLLLLL{to deteriorate}, it is necessary to select appropriate supportive techniques to stabilize the simulations.

The main purpose of this study is to present a scheme for \HLLLLL{the} finite particle approximation of the two-fluid model. We propose to discretize the two-fluid model of superfluid $\rm ^{4}He$ using smoothed particle hydrodynamics (SPH)~\cite{gingold1977smoothed}, a well-established Lagrangian particle approximation \HLLLLL{that is} particularly popular in the field of astrophysics. 
By combining theoretical methodologies typical of two different branches of physics (low-temperature physics and astrophysics), we aim to pioneer a new paradigm of Lagrangian particle mechanics targeting superfluids.

According to quantum mechanics, in principle, \HLLLLL{it is not possible} to separate the superfluid and normal fluid components in space. Let us discuss this point further. \HLLLLL{In recent years, several studies} have combined the vortex filament model (VFM), which describes quantized vortices in superfluid components, with the Navier--Stokes equations, which describe the motion of normal fluids~\cite{Idowu2001, PhysRevLett.120.155301, PhysRevLett.124.155301}. On the premise that this combined approach is acceptable, we can assume that coupling both components of the two-fluid model, instead of using the VFM for the superfluid component, \HLLLLL{would} approximate the system. This approximation \HLLLLL{is} allowed only when we use the two-fluid model to directly simulate large-scale problems \HLLLLL{on the continuum mechanical scale} because technically, this breaks the microscopic laws of quantum mechanics. Scenarios \HLLLLL{such as these}, which simultaneously solve for the two components \HLLLLL{of the two-fluid model}, correspond to multi-phase flows in a classical fluid. \HLLLLL{Therefore}, we allow \HLLLLL{the} two components \HLLLLL{to} be combined in space in the numerical tests in this \HLLLLL{study}. 
\HLLLLL{We investigate the possibility of reproducing several quantum mechanical phenomena observed in liquid $\rm ^{4}He$ in this classical mechanical approximation.}

The remainder of this \HLLLLL{paper} is structured as follows. 
Section~\ref{seq:preperation} describes the concepts of \HLLLLL{the two-fluid model} and SPH, provides a summary of \HLLLLL{the} wave equations in superfluid $\rm ^{4}He$, 
		and discusses the expression of thermodynamic parameters in the two-fluid model, in preparation for Section~\ref{eq:discstandSPH}.
In Section~\ref{eq:discstandSPH}, we derive a finite particle approximation of liquid $\rm ^{4}He$.
%Section~4 examines the discretized parameters of the derived approximation model.
%Section~5 summarizes our results and concludes the paper.
In Section~\ref{sec:strategy}, we discuss strategies for practical applications and introduce several improved numerical schemes for SPH. \HLLLLL{Specifically, we present a reformulation of the viscosity term in the two-fluid model to conserve the angular momentum of the fluid particles around their axes.} In Section~\ref{seq:numericans}, we demonstrate our model with three major numerical tests: the Rayleigh--Taylor instability, wave propagation analyses, and rotating cylinder simulations. Section~\ref{sec:conc} summarizes our results and concludes the paper.

\section{Preparations} \label{seq:preperation}
\subsection{\HLLLLL{Two-fluid model}}\label{seq:twofluidmodel}
The total mass density $\rho$ is expressed as follows:
\begin{eqnarray}
\rho &=& \rho_{n} + \rho_{s}, \label{eq:superpos}
\end{eqnarray}
where $\rho_{n}$ and $\rho_{s}$ are the mass densities of the normal fluid and of the superfluid, which respectively satisfy the laws of mass and entropy conservations as follows: 
\begin{eqnarray}
\frac{\partial \rho}{\partial t} + \vec{\nabla} \cdot (\rho_{n}\vec{v}_{n} + \rho_{s}\vec{v}_{s}) &=& 0, \label{eq:conservmass} \\
\frac{\partial }{\partial t}(\rho \sigma) + \vec{\nabla} \cdot (\rho\sigma\vec{v}_{n}) &=& 0. \label{eq:conserventropy} 
\end{eqnarray}
Here, $\vec{v}_{n}$ and $\vec{v}_{s}$ are the velocities of the normal fluid and of the superfluid, respectively, while $\sigma$ is the entropy density. 
% In the left-hand side of Eq.~(\ref{eq:conserventropy}), the contribution of $\vec{v}_s$ is missing in the bracket of the second term because the superfluid has no entropy.

By solving the Gross-Pitaevskii equation (a nonlinear $\rm Schr\ddot{o}dinger$ equation for boson particles) and the Gibbs-Duhem equation simultaneously, we obtain the following equations of motion for liquid $\rm ^{4}He$ from a phenomenological perspective~\cite{Tsubota2017}:
\begin{eqnarray}
\rho_{s} \frac{{\rm D} \vec{v}_{s}}{{\rm D} t} &=& -\frac{\rho_{s}}{\rho}\nabla P + \rho_{s}\sigma\nabla T \label{eq:goveqsuper}, \\
\rho_{n} \frac{{\rm D} \vec{v}_{n}}{{\rm D} t} &=& -\frac{\rho_{n}}{\rho}\nabla P - \rho_{s}\sigma\nabla T + \eta_{n}\nabla^2 \vec{v}_{n}, \label{eq:goveqnormal}
\end{eqnarray}
where $D\{\cdot\}/Dt$ represents a material derivative, 
and $\vec{v}_{n}$ and $\vec{v}_{s}$ are the velocities of the normal fluid and superfluid, respectively. 
$\eta_{n}$ is the viscosity of the normal fluid. 
$T$ is temperature, $P$ is pressure, and $\sigma$ is entropy density.  

Hydrodynamic aspects of liquid $\rm ^{4} He$ have been intensively studied through experiments involving counterflow in a cylinder attached to a reservoir of the liquid at one point and a heater at another point. Under this condition, the liquid $\rm ^{4} He$ is in a separate phase of super and normal fluids, which flow in the opposite direction~\cite{TOUGH1982133}. C. J. Gorter and J. H. Mellink found the mutual friction forces to be the reason for the discrepancy in Eq.~(\ref{eq:goveqsuper}) and Eq.~(\ref{eq:goveqnormal}) with respect to the counterflow experiments at high heat fluxes~\cite{GORTER1949285}, which are given as
\begin{eqnarray}
\rho_{s} \frac{{\rm D} \vec{v}_{s}}{{\rm D} t} &=& -\frac{\rho_{s}}{\rho}\nabla P + \rho_{s}\sigma\nabla T - \vec{F}_{sn}, \label{eq:goveqsuper:mut}\\
\rho_{n} \frac{{\rm D} \vec{v}_{n}}{{\rm D} t} &=& -\frac{\rho_{n}}{\rho}\nabla P - \rho_{s}\sigma\nabla T + \eta_{n}\nabla^2 \vec{v}_{n} + \vec{F}_{sn}. \label{eq:goveqnormal:mut}
\end{eqnarray}
Further studies by W.F.~Vinen~\cite{doi:10.1098/rspa.1957.0071} through experiments with the second sound wave attenuations revealed a detailed quantitative expression for $\vec{F}_{sn}$ as
\begin{eqnarray}
\vec{F}_{sn} = \frac{2}{3} \rho_{s} \alpha \kappa L \vec{v}_{sn}, \label{eq:mutfricfs}
\end{eqnarray}
%where $\rho_{s}$ is the mass density of a superfluid, $\kappa$ is the quantum of circulation, $\vec{v}_{sn}$ is the relative velocity of $\vec{v}_{s} - \vec{v}_{n}$, $\alpha$ is a friction coefficient, and $L$ is the vortex line density. It is known that $L$ at the steady state can be described as} 
where $\rho_{s}$ is the mass density of a superfluid, $\kappa$ is the quantum of circulation, $\vec{v}_{sn}$ is the relative velocity of $\vec{v}_{s} - \vec{v}_{n}$, $\alpha$ is a friction coefficient, and $L$ is the vortex line density which is time dependent. It is known that the decay of $L$ can be described as 
\HL{
\begin{eqnarray}
L^{-1}(t) = L^{-1}(0) + \beta_{v} t, \label{eq:vineneq}
\end{eqnarray}
where $\beta_{v}$ is a coefficient in the Vinen equation~\cite{NEMIROVSKII201385}.} 
%For more details of $\beta_{v}$, refer to Section~\ref{sec:variable}. }
% For further breakdown of $\beta_{v}$, refer to the literature~\cite{NEMIROVSKII201385}.}

\subsection{A brief overview of standard SPH}
The fundamental concept of SPH is the approximation of the Dirac delta function $\delta$ in integral form by a distribution function $W$, which is called the smoothed kernel function, as follows~\cite{gingold1977smoothed}:
\begin{eqnarray}
\phi(\vec{r}) &=& \int_{\Omega} \phi(\acute{\vec{r}}) \delta(\vec{r} - \acute{\vec{r}})d\acute{\vec{r}} \nonumber \\ 
&\simeq& \int_{\Omega} \phi(\acute{\vec{r}}) W(\vec{r} - \acute{\vec{r}}, h)d\acute{\vec{r}}, \label{eq:approxkernel}
\end{eqnarray}
where $\phi$ is a physical value at the position $\vec{r}$ in a domain $\Omega$ and the parameter $h$ is called a kernel radius. 
\begin{figure}[t]
\vspace{-2.2cm}
%\begin{center}
\includegraphics[width=0.65\textwidth, clip, bb= 0 0 960 720]{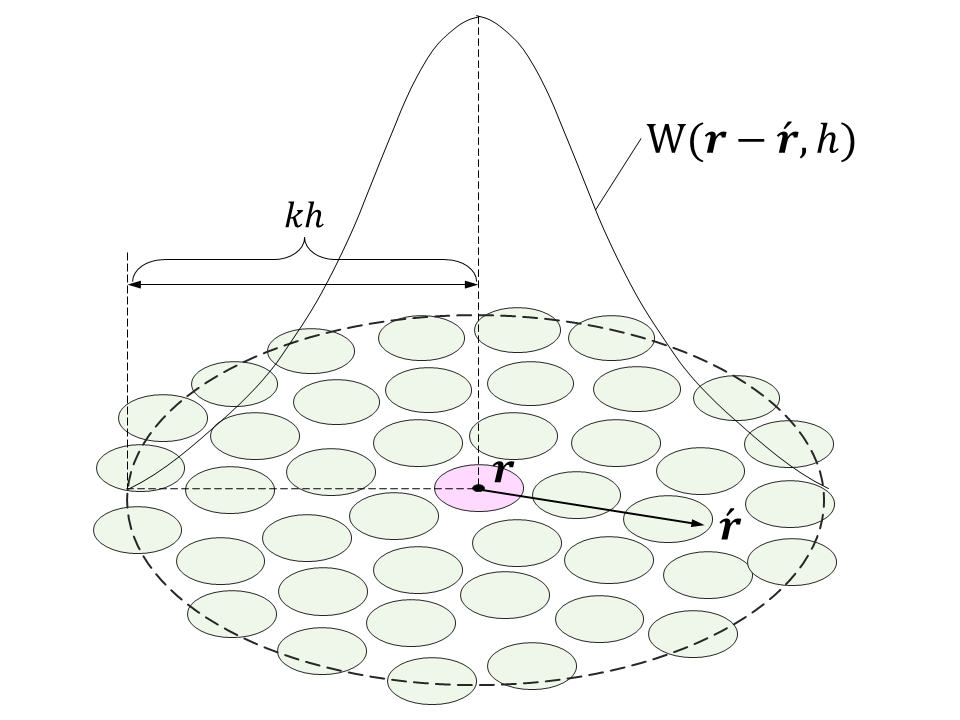}
%\centerline{\includegraphics[scale=0.3]{Figure_KernelFunction.eps}}
%\end{center}
\caption{Schematic of a kernel function.}
\label{fig:SchemKernelFunc}
\end{figure}
The smoothed kernel function $W$ must have at least the following four characteristics; (a) $W$ converges to the delta function $\delta$ as $h$ approaches $0$~($\lim_{h \rightarrow 0} W = \delta$); (b) $W$ satisfies the normalization condition~($\int W d\vec{r}=1$); (c) $W$ satisfies the symmetricity $W(-\vec{r}) = W(\vec{r})$; and (d) $W$ is such that the value of $W$ becomes $0$ at a distance $kh$, which is called a compact support condition.
Figure~\ref{fig:SchemKernelFunc} represents a schematic view of a kernel function. 
An example of a smoothed kernel function $W$ is the Gaussian, which is simply expressed as
\begin{eqnarray}
~W(\vec{r})&:=& \frac{C}{h^{d}}{\rm e}^{-(\vec{r} - \acute{\vec{r}})^2/h^{2}}, \label{eq:gauss}
\end{eqnarray}
where $d$ is the dimension $(d=1,2,~{\rm or}~3)$ and $C$ is a normalization factor.
The Gaussian kernel is frequently used in the field of astrophysics. In the field of incompressible SPH, in contrast, polynomial functions are often used~\cite{desbrun1996smoothed, muller2003particle}. 

Equation~(\ref{eq:approxkernel}) is further discretized on the basis of the summation approximation using the position $\vec{r}_j$, the infinitesimal volume $\Delta V_{j}$, density $\rho_{j}$, and mass $m_{j}$ of the $j$th particle as follows.
\begin{eqnarray}
\phi(\vec{r}_{i}) &\approx& \sum^{N_{p}}_{j}\phi(\vec{r}_{j}) W(|\vec{r}_{i} - \vec{r}_{j}|, h) \Delta V_{j} \nonumber\\
&=& \sum^{N_{p}}_{j}\frac{\phi(\vec{r}_{j})}{\rho_{j}} m_{j} W(|\vec{r}_{i} - \vec{r}_{j}|, h),\label{eq:densityformula}
\end{eqnarray}
where $m_{j} = \rho_{j}\Delta V_{j}$ and $N_{p}$ is the number of fluid particles used to approximate the system.
The gradient and the Laplacian of the physical value $\phi(\vec{r}_{i})$ are axiomatically obtained from Eq.~(\ref{eq:densityformula}) using a vector analysis, as follows:
\begin{eqnarray}
\nabla \phi(\vec{r}_{i}) 
	&=& \rho_{i}\sum^{N_{p}}_{j} m_{j} 
	\Bigl(\frac{\phi(\vec{r}_{i})}{\rho_{i}^2} + \frac{\phi(\vec{r}_{j})}{\rho_{j}^2}\Bigr) 
			\nabla W_{ij},\label{eq:gradient} \\
\nabla^{2} \phi(\vec{r}_{i}) 
	&=& \sum^{N_{p}}_{j} \frac{m_{j}}{\rho_{j}}\frac{\phi(\vec{r}_{i})-\phi(\vec{r}_{j})}{|\vec{r}_{i}-\vec{r}_{j}|^{2}}  
			(\vec{r}_{i} - \vec{r}_{j})\cdot \nabla W_{ij},\label{eq:laplacian}
\end{eqnarray}
where $W_{ij} = W(|\vec{r}_{i} - \vec{r}_{j}|, h)$.
Here, we have used the mathematical relationship $\nabla f = \rho[\nabla(f/\rho)+(f/\rho^2)\nabla \rho]$~to derive Eq.~(\ref{eq:gradient}), so that the gradient operator has symmetricity of pressures between the $i$th and $j$th particles. 
For more details of the operators in standard SPH, refer to~\cite{doi:10.1146/annurev.aa.30.090192.002551}. 

\subsection{Wave equations of superfluids}
According to the literature~\cite{PhysRev.60.356}, the governing equations of Eq.~(\ref{eq:conservmass}) to Eq.~(\ref{eq:goveqnormal}) yield wave propagation equations regarding density $\rho$ and entropy $S$, which produce multiple types of sound waves. Under a first approximation that neglects the viscosity term, the wave equations give two different sound waves, expressed as follows~\cite{donnelly2009two, LIFSHITZ1992177}:
\begin{eqnarray}
c_{1} &=& \biggl(\frac{\partial P}{\partial \rho}\biggr)_{S}^{1/2},\\
c_{2} &=& \biggl(\frac{\rho_{s}}{\rho_{n}}\frac{TS^2}{C_{v}}\biggr)^{1/2}, \label{eq:secondwave}
\end{eqnarray}
where $c_{1}$ and $c_{2}$ are called the first sound velocity and the second sound velocity, respectively. 
$C_{v}$ represents the specific heat at constant volume.
Equation~(\ref{eq:secondwave}) can be rewritten using the number of $\rm ^{4}He$ atoms $N$ and the mass of a $\rm ^{4}He$ atom $M$ as
\begin{eqnarray}
\frac{\rho_{s}}{\rho_{n}} = \Biggl\{\frac{C_{v}c_{2}^2}{(NM)^{2} T \sigma^{2}}\Biggr\}, \label{eq:ratiorho}
\end{eqnarray}
where we have introduced a basic relation between the entropy density $\sigma$ and the entropy $S$ of $\sigma=S/(NM)$.

\subsection{Expression of thermodynamic variables in an elementary excitation model}
L.~Landau derived that the elementary excitation of liquid $\rm ^{4}He$ consists of two components, ``phonons'' and ``rotons''~\cite{PhysRev.60.356}.
The pressure $P$ of liquid $\rm ^{4}He$ can be expressed as follows~\cite{Adamenko_2008}:
\begin{eqnarray}
P &=& P_{\rm ph} + P_{\rm rot}. \label{eq:Pphrot} 
\end{eqnarray}
Here, let us denote the ratio of a circle's circumference as $\pi$, and the speed of sound as $c$.
We also introduce constant values for ~$\mu$, $p_{0}$, and $\Delta$ as ${\rm 1.72\times10^{-24}~g}$, ${\rm 2.1\times10^{-19}~gcms^{-1}}$, and ${\rm 8.9~K}$, respectively, according to the literature~\cite{schmitt2015introduction, bennemann2013novel}.
When $T \ll {\rm 93~K}$, Eq.~(\ref{eq:Pphrot}) can be approximated as~\cite{Adamenko_2008, schmitt2015introduction}:
\begin{eqnarray}
P &\simeq& \frac{\pi^2 (k_{B}T)^{4}}{90 \hbar^3 c^{3}} + \frac{p_{0}^{2}\sqrt(\pi\mu/2)}{2\pi^2 \hbar^3}(k_{B}T)^{3/2} e^{-\Delta/T}, \label{eq:goldpres}
\end{eqnarray}
where $k_{B}$ represents the Boltzmann constant and $\hbar$ is the reduced Planck's constant.
Note that Eq.~(\ref{eq:goldpres}) is similar to  Eq.~(5) in the literature~\cite{Adamenko_2008}. 
The entropy $S$ is obtained from the first-order partial derivatives of $P$ with respect to temperature $T$ as
\begin{eqnarray}
S &:=& \biggl(\frac{\partial P}{\partial T}\biggr) \\
&\simeq& \frac{2\pi^2 k_{B}^4 T^{3}}{45 \hbar^3 c^{3}} + \biggl(\frac{k_{B}}{2\pi}\biggr)^{3/2}\frac{\sqrt{\mu}p_{0}^2 \Delta}{\hbar^3}\frac{e^{-\Delta/T}}{\sqrt{T}}, \label{eq:goldentro}
\end{eqnarray}
where we have ignored higher-order terms in the derivation of the second term in Eq.~(\ref{eq:goldentro}).

Accordingly, the entropy density $\sigma$ is obtained using the relation $\sigma=S/(NM)$ as
\begin{eqnarray}
\sigma &=& \zeta T^{3} + \xi \frac{e^{-\Delta/T}}{T^{1/2}}, \label{eq:entrodens} \\
\zeta &:=& \frac{2\pi^2 k_{B}^4}{45 \hbar^3 c^{3}NM}, \label{eq:entrodens2} \\ 
\xi &:=& \biggl(\frac{k_{B}}{2\pi}\biggr)^{3/2}\frac{\sqrt{\mu}p_{0}^2 \Delta}{\hbar^3 NM}. 
\end{eqnarray}
Here, both $\zeta$ and $\xi$ are constant parameters because their components of $N, M, \pi, c, \mu, p_{0}, \Delta, k_{B},~{\rm and~} \hbar$ are all constant and can be given as known parameters.
Now, we have prepared all the parameters needed for the derivation of a particle approximation of liquid $\rm ^{4}He$.

% \section{A particle approximation for superfluids}
\HL{\section{Discretization of the two-fluid model using standard SPH}\label{eq:discstandSPH}}
First, Eq.~(\ref{eq:goveqsuper:mut}), the motion equation for superfluid components, can be rewritten as:
\begin{eqnarray}
\frac{{\rm D} \vec{v}_{s}}{{\rm D} t} &=& -\frac{1}{\rho}\nabla P + \sigma\nabla T - \HL{\frac{1}{\rho_{s}}\vec{F}_{sn}} \label{eq:goveqsuperdiv}. 
\end{eqnarray}
Using Eq.~(\ref{eq:gradient}), we obtain the particle discretization of the right-hand side of Eq.~(\ref{eq:goveqsuperdiv}), with respect to the $i$th particle, as follows:
\begin{eqnarray}
\Biggl\langle -\frac{1}{\rho}\nabla P\Biggr\rangle_{i} 
&=& -\sum_{j} m_{j} 
	\Bigl(\frac{P_{i}}{\rho_{i}^2} + \frac{P_{j}}{\rho_{j}^2}\Bigr) 
			\nabla W_{ij},\label{eq:pdiscsuperfirst} \\
\Biggl\langle \sigma\nabla T  \Biggr\rangle_{i}
&=&\theta^{s}_{i}\sum_{j} m_{j} 
	\Bigl(\frac{T_{i}}{\rho_{i}^2} + \frac{T_{j}}{\rho_{j}^2}\Bigr) 
			\nabla W_{ij},\label{eq:pdiscsupersecond} 
\end{eqnarray}
where $\theta^{s}_{i}$ is represented  as
%where $\theta^{s}_{i}$ is represented using Eq.~(\ref{eq:entrodens}) as
\begin{eqnarray}
\theta^{s}_{i} &=& \rho_{i}\sigma_{i}. \label{eq:disc:super:theta}
\end{eqnarray}
Here, the superscript of $\theta^{x}_{i}$ represents the first letter of ``superfluid'' or ``normal fluid''~($x$ = s, n).
% $\sigma_{i}$ is a specific case of Eq.~(\ref{eq:entrodens}) with $T=T_{i}$.
\HL{The detailed evaluation of $\sigma_{i}$ is discussed in Section~\ref{sec:calcentro}.}

\HL{
The two-way coupling technique for SPH~\cite{ROBINSON2014121, HE2018548} that uses the weighted average method to compute relative velocity between two phases 
%	, which conserves an equal amount of momentum exchange between the two phases, 
can be applied to the discretization of $\vec{F}_{sn}$. From Eq.~(\ref{eq:mutfricfs}), a mutual friction force $\vec{F}_{s}^{(i)}$ of the $i$th superfluid particle is calculated as follows:}
\HL{
\begin{eqnarray}
\vec{F}_{s}^{(i)} &=& C_{L} \frac{\sum_{j \in \Omega_{i}} (\vec{v}_{s}^{(i)} - \vec{v}_{n}^{(j)}) W_{ij}}{\sum_{j \in \Omega_{i}} W_{ij}}, \label{eq:mutualdisc}
\end{eqnarray}
where $C_{L}$ is $2/3\rho_{s} \alpha \kappa L$ and $\Omega_{i}$ is the set of all the particles in the neighborhood of the $i$th normal fluid particles.}
\HL{The third term on the right-hand side in Eq.~(\ref{eq:goveqsuperdiv}) is obtained by \HLLLLL{multiplying} Eq.~(\ref{eq:mutualdisc}) by $-\rho_{s}^{-1}$. }
% \HL{Let us evaluate the parameter $C_{L}$ in Section~\ref{sec:calcentro}.}

%\HL{The value of $C_{L}$ is estimated in Section~\ref{sec:calcentro}. }

Second, Eq.~(\ref{eq:goveqnormal:mut}), the motion equation for normal fluid components, can be rewritten as:
\begin{eqnarray}
\frac{{\rm D} \vec{v}_{n}}{{\rm D} t} &=& -\frac{1}{\rho}\nabla P - \frac{\rho_{s}}{\rho_{n}}\sigma\nabla T + \frac{\eta_{n}}{\rho_{n}}\nabla^2 \vec{v}_{n} \HL{+ \frac{1}{\rho_{n}}\vec{F}_{sn}}. \label{eq:goveqnormaldiv}
\end{eqnarray}
The first term on the right-hand side in Eq.~(\ref{eq:goveqnormaldiv}) can be discretized similar to the case of Eq.~(\ref{eq:pdiscsuperfirst}). Meanwhile, the discretized expression of the second term with respect to the $i$th particle is obtained using Eq.~(\ref{eq:gradient}) as follows.
\begin{eqnarray}
\Biggl\langle - \frac{\rho_{s}}{\rho_{n}}\sigma\nabla T \Biggr\rangle_{i} 
&=&\theta^{n}_{i}\sum_{j} m_{j} 
	\Bigl(\frac{T_{i}}{\rho_{i}^2} + \frac{T_{j}}{\rho_{j}^2}\Bigr) 
			\nabla W_{ij},\label{eq:pdiscnormalfirst} 
\end{eqnarray}
where the breakdown of $\theta^{n}_{i}$ is given as
\begin{eqnarray}
\theta^{n}_{i} &=& \chi\frac{\rho_{i}}{T_{i}\sigma_{i}}, \label{eq:disc:normal:theta} \\
		\chi &:=& \frac{C_{v}c_{2}^{2}}{(NM)^2}. \label{eq:disc:normal:chi}
\end{eqnarray}
\HL{Meanwhile}, a substitution of Eq.~(\ref{eq:laplacian}) into the third term directly leads to:
\begin{eqnarray}
\Biggl\langle \frac{\eta_{n}}{\rho_{n}}\nabla^2 \vec{v}_{n} \Biggr\rangle_{i} 
= \nu_{n} \sum_{j} \frac{m_{j}}{\rho_{j}}\frac{\vec{v}_{n}^{(i)}-\vec{v}_{n}^{(j)}}{|\vec{r}_{i}-\vec{r}_{j}|^{2}}  
			(\vec{r}_{i} - \vec{r}_{j})\cdot \nabla W_{ij},\nonumber \\
				\label{eq:laplaciandisc}
\end{eqnarray}
where $\nu_{n}$ is the kinetic viscosity, which is equal to~$\eta_{n}/\rho_{n}$.
\HL{Subsequently, the mutual friction force $\vec{F}_{n}^{(j)}$ of the $j$th normal fluid particle is obtained by summing up the contributions from each reaction force of the $i$th superfluid particle as follows:}
\HL{
\begin{eqnarray}
\vec{F}_{n}^{(j)} &=& - \sum_{i \in \Omega_{j}} \vec{F}_{s}^{(i)} \Bigl( \frac{W_{ij}}{\sum_{k \in \Omega_{i}} W_{ik}} \Bigr), \label{eq:mutualdisc:normal}
\end{eqnarray}
where $\Omega_{j}$ is the set of all particles in the neighborhood of the $j$th normal fluid particle, while $\Omega_{i}$ is that of the $i$th superfluid particle.
Note that the summation inside the brackets in Eq.~(\ref{eq:mutualdisc:normal}) is with respect not to $j$ but to $i$ in order for the two phases to conserve an equal amount of momentum exchange.
}
\HL{In the end, the fourth term on the right-hand side in Eq.~(\ref{eq:goveqnormaldiv}) is obtained by \HLLLLL{multiplying} Eq.~(\ref{eq:mutualdisc:normal}) by $\rho_{n}^{-1}$.}

Now, the right-hand sides of Eq.~(\ref{eq:goveqsuperdiv}) and Eq.~(\ref{eq:goveqnormaldiv}) are appropriately discretized. 
The material derivative of velocities in the left-hand sides of Eq.~(\ref{eq:goveqsuperdiv}) and Eq.~(\ref{eq:goveqnormaldiv}) can be discretized using an explicit time-integrating scheme, e.g., the Verlet algorithm~\cite{PhysRev.159.98}, in accordance with the usual manner in SPH simulations. Consequently, we have successfully derived a particle approximation of the governing equations of liquid $\rm ^{4}He$. \\ 

\HL{\section{Proposed strategies for practical application} \label{sec:strategy}}

\HL{\subsection{Introduction of improved techniques to stabilize the simulations}}
\HL{
It is known that the standard SPH only permits a slight density difference between two phases when being applied to multi-phase flow problems because it does not assume a larger density gradient than that of the gradient of the smoothing kernel function in its formulation~\cite{MONAGHAN1995225}. As a remedial measure, we introduce an improved SPH scheme that ensures the continuity of the pressure and pressure gradient at the interface between two phases, while satisfying mass and momentum conservation~\cite{HU2006844}. In this method, the density of the $i$th particle is computed as follows:}
\HL{\begin{eqnarray}
\rho_{i} &=& m_{i}\sum_{j} W(|\vec{r}_{i} - \vec{r}_{j}|, h). \label{eq:Hu:dens}
\end{eqnarray}
Instead of Eq.~(\ref{eq:pdiscsuperfirst}), the pressure gradient term is then computed as}
\HL{
	\begin{align}
	{ \small
	\Biggl\langle -\frac{1}{\rho}\nabla P\Biggr\rangle_{i} 
	= -\frac{1}{m_{i}}\sum_{j}  
	\Bigl(\frac{m_{i}^2}{\rho_{i}^2} + \frac{m_{j}^2}{\rho_{j}^2}\Bigr)\tilde{P}~ 
		\nabla W_{ij}},
		\label{eq:huadam2006:pg} 
	\end{align}
where $\tilde{P}$ is the reduced pressure, given as}
\HL{
	\begin{eqnarray}
	\tilde{P} &:=& \frac{\rho_{i}P_{j} + \rho_{j}P_{i}}{\rho_{i}+\rho_{j}}. 
	\end{eqnarray}
}

\HL{
Additionally, several sophisticated techniques are appropriately introduced. First, we use a pair-wise particle collision technique~\cite{SHAKIBAEINIA201213} that gives a repulsive force when two particles are too close to each other. Specifically, the momentum conservation in the normal direction between two colliding particles is computed using this technique. Next, we introduce a one-dimensional Riemann solver~\cite{ZHANG2017605} in each pair-wise particle, instead of using artificial viscosities. These two supportive techniques become valid only when the two particles are closer to each other than a distance equal to the initial particle distance $d$, as exceptional cases.
}

\HLLLLL{
%The non-uniformity of particle distributions leads to particle shortages at local spaces in a simulation domain, which can deteriorate the numerical accuracy drastically to cause unphysical pressure oscillations. Hence, it is indispensable to utilize stabilization techniques to this end. The related works of~\cite{KondoKoshizuka2011, asai2012stabilized} proposed a new formulation of the pressure Poisson equation with relaxation coefficient that contributes to obtaining smoothed pressure distributions in the implicit or semi-implicit particle methods. In respect of SPH using explicit time-integrating schemes, \cite{antuono2010free} presented a weakly compressible SPH improved by adding artificial diffusive terms to the continuity equations. Several filtering techniques have also been reported; \cite{doi:10.1146/annurev.aa.30.090192.002551} proposed a smoothing velocity by interpolation among neighboring particles. \cite{pannizo2004physical} presented a density filtering technique known as the Shepard filter, a zero-th order correction among neighboring particles. This study adopts a pressure interpolant with a similar formulation as the Shepard filter by reference to the literature~\cite{imoto2019convergence}.
The non-uniformity of particle distributions leads to particle shortages in the local spaces in the simulation domain, which can cause the numerical accuracy to drastically deteriorate and cause unphysical pressure oscillations. Hence, the utilization of stabilization techniques to this end is indispensable. A new formulation of the pressure Poisson equation with a relaxation coefficient that contributes to obtaining smoothed pressure distributions in the implicit or semi-implicit SPH was proposed~\cite{asai2012stabilized}. With respect to the use of explicit time-integrating schemes in the SPH method, a weakly compressible SPH improved by adding artificial diffusive terms to the continuity equations was presented~\cite{antuono2010free}. Several filtering techniques have also been reported. For example, a smoothing velocity involving interpolation among neighboring particles was proposed~\cite{doi:10.1146/annurev.aa.30.090192.002551}. A density filtering technique known as the Shepard filter for the zero-th order correction among neighboring particles was developed~\cite{pannizo2004physical}. This study adopts a pressure interpolant with a similar formulation as the Shepard filter, according to the literature~\cite{imoto2019convergence}.
}

\HLLLLL{
%In the case of classical fluids, we usually consider introducing surface tension models~\cite{BRACKBILL1992335, LAFAURIE1994134, Abdelraheem2013} to describe the mechanics between two phases accurately. However, it is safer to avoid using existing surface tension models in our case because these previous models build on the Young-Laplace relationship in continuum mechanics and assume the dynamics of classical turbulence. Above all, it is known that the liquid $\rm ^{4}He$ exhibits quantum turbulence~\cite{Golov2009, PhysRevB.97.184518}. Although several recent studies report that both quantum and classical turbulent systems show common characteristics~\cite{Lvov2008, PhysRevB.84.054525}, e.g., the Kolmogorov-1941 energy spectrum~\cite{1941DoSSR..30..301K}, the detailed correspondence between the two different turbulent systems are still in discussion. For these reasons, we avoid using conventional surface tension models.
In the case of classical fluids, we usually consider introducing surface tension models~\cite{BRACKBILL1992335, LAFAURIE1994134, Abdelraheem2013} to describe the mechanics between two phases accurately. However, it is safer to avoid using existing surface tension models in our case because these previous models build on the Young--Laplace relationship in continuum mechanics and assume the dynamics of classical turbulence. Above all, it is known that liquid $\rm ^{4}He$ exhibits quantum turbulence~\cite{Golov2009, PhysRevB.97.184518}. Although several recent studies report that quantum and classical turbulent systems both have common characteristics~\cite{Lvov2008, PhysRevB.84.054525}, for example, the Kolmogorov1941 energy spectrum~\cite{1941DoSSR..30..301K}, details of the correspondence between the two different turbulent systems are still in discussion. Therefore, we avoid using conventional surface tension models.}

Meanwhile, we set the time increment, $\delta t$, to be smaller than the three conditions of numerical stability: the Courant--Friedrichs--Lewy (CFL) condition, the diffusion condition, and the body force condition~\cite{XU201643}. We use an explicit time-integrating scheme, the \HLLLLL{velocity-}Verlet algorithm~\cite{PhysRev.159.98, 10.5555/76990}, in accordance with the usual process for SPH simulations. It should be noted that there exists a possibility to violate stringent energy conservation when using an explicit time-integrating scheme. Regarding entropy conservation, we set constant values for the particles in the initial state and retained these values during simulations; this is further explained in \HLLLLL{Section~\ref{sec:calcentro}}.\\\\

\HLLLLL{\subsection{A reformulation of the viscosity term in the two-fluid model} \label{sec:extension}}
The viscosity term in Eq.~(\ref{eq:laplaciandisc}) also requires improvement because this study focuses on the problem of liquid $\rm ^{4}He$ rotated by a cylindrical vessel along a single direction at a constant angular velocity. In this case, the fluid particles' angular momentum around their axes (often called ``spin angular momentum'' as an analogy for the corresponding term in quantum mechanics) could contribute to the dynamics of the entire system.  However, many SPH series, including those reported in the literature~\cite{HU2006844}, do not consider the spin angular momentum of fluid particles. This is mainly because these previous works focus on governing equations that omit the rotational contribution by the spin angular momentum of particles. Although neglecting rotational terms does not cause a problem in many cases, it can result in severe issues in certain cases, particularly for a fluid containing insolvable matters such as vesicle suspensions.

In 1964, D. W. Condiff derived the Navier--Stokes equations in the Lagrangian form with spin angular momentum conservation from Cauchy's linear momentum principle, aiming to reproduce the tiny effect of rotating molecules~\cite{doi:10.1063/1.1711295}. The resulting equation is expressed as
\begin{eqnarray}
		\rho \frac{d\vec{v}}{dt} &=& -\nabla P + (\bar{\eta} + \bar{\eta_{r}})\nabla^2 \vec{v}, \nonumber \\ 
		&+& \biggl( \frac{\bar{\eta}}{3} + \bar{\xi} -\bar{\eta_{r}} \biggr) \nabla\nabla\cdot\vec{v} + 2\bar{\eta_{r}}\nabla\times\vec{\omega}. \label{eq:NSwithangvconsv}
\end{eqnarray}
Here, $\bar{\eta}$, $\bar{\eta_{r}}$, and $\bar{\xi}$ are the shear viscosity, rotational viscosity, and bulk viscosity, respectively. $\bar{\eta_{r}}$ determines the magnitude of rotational forces and must satisfy the relationship ${\rm 0} \le \bar{\eta_{r}} \le \bar{\eta}$~\cite{MULLER2015301}. The third term on the right-hand side becomes 0 in the case of an incompressible SPH because of the condition $\nabla \cdot \vec{v} = 0$. 

It can be understood that Eq. (\ref{eq:NSwithangvconsv}) is an extension of the ordinary Navier--Stokes equations; the terms on the right-hand side, except for the first term, converge to $\bar{\eta}\nabla^2 \vec{v}$ as the parameter $\bar{\eta_{r}}$ becomes 0 when $\nabla \cdot \vec{v} = 0$. In addition, K. M{\"{u}}ller  discretized Eq.~(\ref{eq:NSwithangvconsv}) using standard SPH in his study on formulating angular momentum conservation in smoothed dissipative particle dynamics~\cite{MULLER2015301}. Given these previous studies, we propose introducing the same extension as Eq.~(\ref{eq:NSwithangvconsv}). Namely, we extend $\eta_{n} \nabla^2 \vec{v}$ in Eq.~(\ref{eq:laplaciandisc}) to the sum of $(\bar{\eta} + \bar{\eta_{r}})\nabla^2 \vec{v}$ and $2\bar{\eta_{r}}\nabla\times\vec{\omega}$, which implies that $\bar{\eta} = \eta_{n}$. We then compute $(\bar{\eta} + \bar{\eta_{r}})\nabla^2 \vec{v}$ using Eq.~(\ref{eq:laplacian}) and $2\bar{\eta_{r}}\nabla\times\vec{\omega}$ for the $i$th particle using K. M{\"{u}}ller's model~\cite{MULLER2015301}, as follows:
\begin{eqnarray}
(\nabla \times \vec{\omega})_{i} &=& \sum_{j} \frac{m_{j}}{\rho_{j}}\nabla W_{ij} \times (\vec{\omega}_{i} + \vec{\omega}_{j}), \label{eq:rotforce}
\end{eqnarray}
where $\vec{\omega}_{i}$ and $\vec{\omega}_{j}$ are the angular velocities of the $i$th and $j$th particles, respectively. 
We set the $\vec{\omega}$ of each particle to $C_\omega \vec{v}_{\rm max}/0.5d$ in the initial state, where $d$ indicates the initial particle distance, $\vec{v}_{\rm max}$ represents the estimated maximum velocity, and $C_{\omega}$ is a constant parameter determining the scale ratio to $\vec{v}_{\rm max}$. The value of $C_{\omega}$ lies between $\rm 0$ and $\rm 1.0$, depending on the target problems in the simulation tests.\\
%Similarly, the following equation is used to compute the vicosity term instead of Eq.~(\ref{eq:laplaciandisc}):}
%\HL{
%	\begin{eqnarray}
%	\begin{align}
%	{ \small
%	\Biggl\langle \frac{\eta_{n}}{\rho_{n}}\nabla^2 \vec{v}_{n} \Biggr\rangle_{i} 
%	= \frac{\eta_{n}}{m_{i}} \sum_{j} \Bigl(\frac{m_{i}^2}{\rho_{i}^2} + \frac{m_{j}^2}{\rho_{j}^2}\Bigr)
%		\frac{\vec{v}_{n}^{(ij)}}{|\vec{r}_{ij}|^{2}} 
%		\vec{r}_{ij}\cdot \nabla W_{ij}}, 
%			%\nonumber \\
%		\label{eq:huadam2006:visc} 
%			\end{align}
%		\end{eqnarray}
%where $\vec{v}_{n}^{(ij)}$ and $\vec{r}_{ij}$ are respectively given as}
%\HL{
%	\begin{eqnarray}
%		\vec{v}_{n}^{(ij)}&:=&\vec{v}_{n}^{(i)} - \vec{v}_{n}^{(j)},\nonumber \\
%		\vec{r}_{ij}&:=&\vec{r}_{i} - \vec{r}_{j}.
%	\end{eqnarray}
%}

\subsection{Entropy density estimation} \label{sec:calcentro}
\begin{figure}[t]
\vspace{-1.8cm}
%\begin{center}
\includegraphics[width=0.48\textwidth, clip, bb= 0 0 960 720]{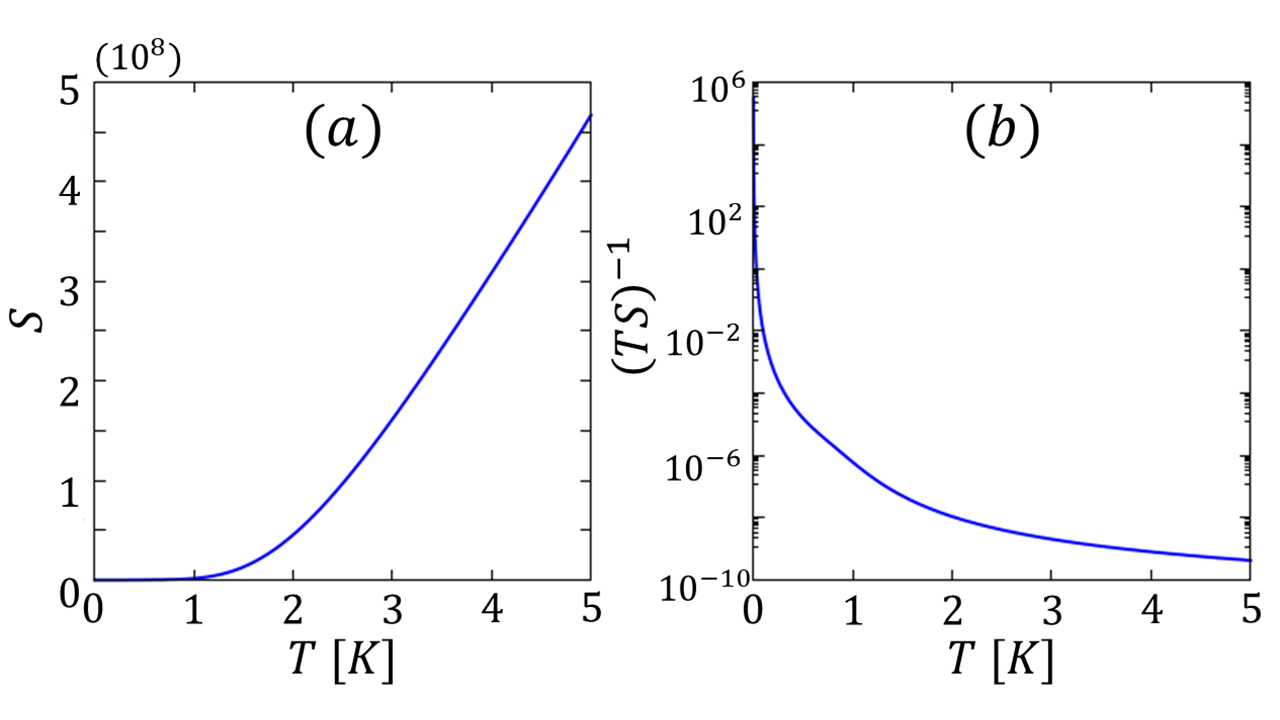}
%\centerline{\includegraphics[scale=0.25]{Figure_ThetaAmp.eps}}
%\end{center}
\caption{Dependences of (a) the entropy $S$ and (b) $(TS)^{-1}$ on the temperature $T$.}
\label{fig:ThetaAmp}
\end{figure}

% Let us discuss the mechanical effect of the discretized parameters of $\theta_{i}^{s}$ and $\theta_{i}^{n}$. 
It can be said from Eq.~(\ref{eq:pdiscsupersecond}) and Eq.~(\ref{eq:pdiscnormalfirst}) that each of $\theta_{i}^{s}$ and $\theta_{i}^{n}$ functions as a parameter that determines the degree of the effect of temperature gradient forces.
The entropy density $\sigma_{i}$ is dominant in the behavior of $\theta_{i}^{s}$ because the density $\rho_{i}$ fluctuates typically less than one percent 
% and at most a few percent 
compared with its averaged density in incompressible SPH~\cite{MONAGHAN1994399, NOMERITAE2016156}. Additionally, the behavior of $\sigma$ corresponds to that of entropy $S$ because the value of $NM$ is constant; thus, $\theta_{i}^{s}$ is proportional to the entropy $S$. 
%Figure~\ref{fig:ThetaAmp}(a) shows the dependence of entropy $S$ on temperature $T$. 
%The following fact is suggested from Fig.~\ref{fig:ThetaAmp}(a);~The effect of the temperature gradient force controlled by $\theta_{i}^{s}$ on the superfluid component diminishes as the temperature $T$ decreases, and it eventually reaches zero at absolute zero.  
Figure~\ref{fig:ThetaAmp}(a) shows the dependence of entropy $S$ on temperature $T$, indicating that the effect of the temperature gradient force controlled by $\theta_{i}^{s}$ on the superfluid component diminishes as the temperature $T$ decreases, and it eventually reaches zero at absolute zero.
Based on the result, we can say that the parameter $\theta_{i}^{s}$ reflects well the characteristic that superfluids have almost zero or a very small temperature gradient, corresponding to extremely high heat conductivity.
%~\cite{PhysRev.162.143}.

\HL{Regarding the parameter $\theta_{i}^{n}$,} the second sound velocity $c_{2}$ \HL{of Eq.~(\ref{eq:disc:normal:chi})} should be semi-empirically determined according to the literature, exemplified by~\cite{PhysRevB.14.3868, wilks1987introduction}. 
\HL{In experiments,} it is known that $c_{2}$ emerges when the system reaches temperatures lower than the critical temperature of approximately $\rm 2.17~K$, increases as the cooling of the system increases, and finally reaches approximately $\rm 135~ms^{-1}$ after experiencing a sluggishness \HL{at approximately $\rm 20~ms^{-1}$}~\cite{wilks1987introduction}. 
Let us assume in this discussion that $c_{2}$ is constant. In this ideal case, the parameter $\chi$ becomes constant; thus, $\theta_{i}^{n}$ becomes proportional to $(T_{i}\sigma_{i})^{-1}$. Also, the behavior of $\sigma$ corresponds to that of entropy $S$, similar to the case of $\theta_{i}^{s}$. After all, the behavior of $\theta_{i}^{n}$ is determined by that of $(TS)^{-1}$. Figure~\ref{fig:ThetaAmp}(b) shows the dependence of $(TS)^{-1}$ on the temperature $T$ with a semi-logarithmic scale. The effect of $\theta_{i}^{n}$ on fluid $\rm ^{4}He$ is confirmed to greatly increase as the temperature $T_{i}$ decreases. 
\HL{Remember that} the parameter $\theta_{i}^{n}$ is the discretized formula of Eq.~(\ref{eq:ratiorho}), expressing the ratio of superfluid density to normal density, which generates the second sound wave.
Figure~\ref{fig:ThetaAmp}(b) suggests that the effect of $\rho_{s}/\rho_{n}$ increases as the temperature $T$ decreases, on the premise that $c_{2}$ is constant; the effect is more amplified in actual cases because the velocity $c_{2}$ itself increases, as mentioned above~\cite{wilks1987introduction}. 

%In any case, both $\theta_{i}^{s}$ and $\theta_{i}^{n}$ are confirmed to reflect the phenomenological behavior of superfluid $\rm ^{4}He$ qualitatively.
\HL{As discussed above, both parameters $\theta_{i}^{s}$ and $\theta_{i}^{n}$ are confirmed to reflect the phenomenological behavior of superfluid $\rm ^{4} He$ qualitatively. 
However, it is also found that $\theta_{i}^{n}$ is too sensitive to be used for practical applications. As a remedy, we give the parameters of $\rho_s$ and $\rho_n$ as constant initial parameters; we compute  Eq.~(\ref{eq:pdiscnormalfirst}) from Eq.~(\ref{eq:pdiscsupersecond}) multiplied by $-\rho_{s}/\rho_{n}$. We also compute the entropy $\sigma_{i}$ in the parameter $\theta_{i}^{s}$ from Eq.~(\ref{eq:entrodens}) scaled by a coefficient of $C_{e}$ as} 
\HL{\begin{eqnarray}
\sigma_{i} &=& C_{e} \sigma. \label{eq:entrodens3}
\end{eqnarray}}
Here, the reason we introduce $C_{e}$ is because Eq.~(\ref{eq:goldentro}) does not consider the effect of interactions among particles in its derivation. In that sense, Eq.~(\ref{eq:goldentro}) can be said to express a relationship between entropy and temperature in a quantum ideal gas. Hence, it is necessary to consider the volume ratio of gas to a liquid. In this article, we estimate the value of $C_{e}$ as the volume ratio of gas to liquid, each of which has the same density under a certain pressure ($C_{e} \approx V_{\rm liquid}/V_{\rm air}$), as the first approximation level.

\subsection{The variables of the system}\label{sec:variable}
The variables for the $i$th particle are given as a set of $(\vec{r}_{i}, \vec{v}_{n}^{(i)}, \vec{v}_{s}^{(i)}, \vec{\omega}_{i}, m_{i}, \rho_{i}, \sigma_{i}, P_{i}, T_{i})$. $P_{i}$ is calculated from $\rho_{i}$ by solving a one-to-one thermodynamic correspondence between $P_{i}$ and $\rho_{i}$ given by the Tait equation of state~\cite{Monaghan_2005}. \HL{Meanwhile, $\rho_{i}$ and $\sigma_{i}$ are calculated using Eq.~(\ref{eq:Hu:dens}) and  Eq.~(\ref{eq:entrodens3}), respectively. $m_{i}$ and $T_{i}$ keep their initial values in this paper.}
\HL{$\vec{v}_{s}^{(i)}$ and $\vec{v}_{n}^{(i)}$ are not necessarily distinguished in implementations because each particle only takes either a superfluid or normal fluid state at a time. Thus, the variable of the $i$th particle can be reduced to the set of $(\vec{r}_{i}, \vec{v}_{i}, \vec{\omega}_{i}, m_{i}, \rho_{i}, \sigma_{i}, P_{i}, T_{i})$.}

\HL{The parameter $C_{L}$ includes one time-dependent parameter of the vortex line density $L$ and three other constant values: the superfluid density $\rho_s$, a friction coefficient $\alpha$, and the quantum of circulation $\kappa$, where $L$ is calculated from Eq.~(\ref{eq:vineneq}) at every time step. In this article, the parameter $\beta_{v}$ of Eq.~(\ref{eq:vineneq}) is calculated according to Table~1 and Table~3 in the literature~\cite{Vinen2002}.} 
\HL{To summarize,} the constant parameters characterizing the system are given as a set of $(N, M, \HL{\rho_{s}}, \HL{\rho_{n}}, \HL{T_{0}}, \HL{C_{e}}, \HL{\bar{\eta}}, \HL{\bar{\eta}_{r}}, c, \HL{\alpha, \HL{\beta_{v}}, \kappa}, \HL{L_{0}})$, \HL{where $L_{0} = L(0)$ of Eq.~(\ref{eq:vineneq})}. The exact constant values that are independent of the system are $(\mu, p_{0}, \Delta, k_{B}, \hbar)$.

\subsection{Validation of SPH model}\label{sec:modelvtests}
\begin{figure}[t]
\vspace{-3.5cm}
%\begin{center}
\includegraphics[width=0.64\textwidth, clip, bb= 0 0 1280 1440]{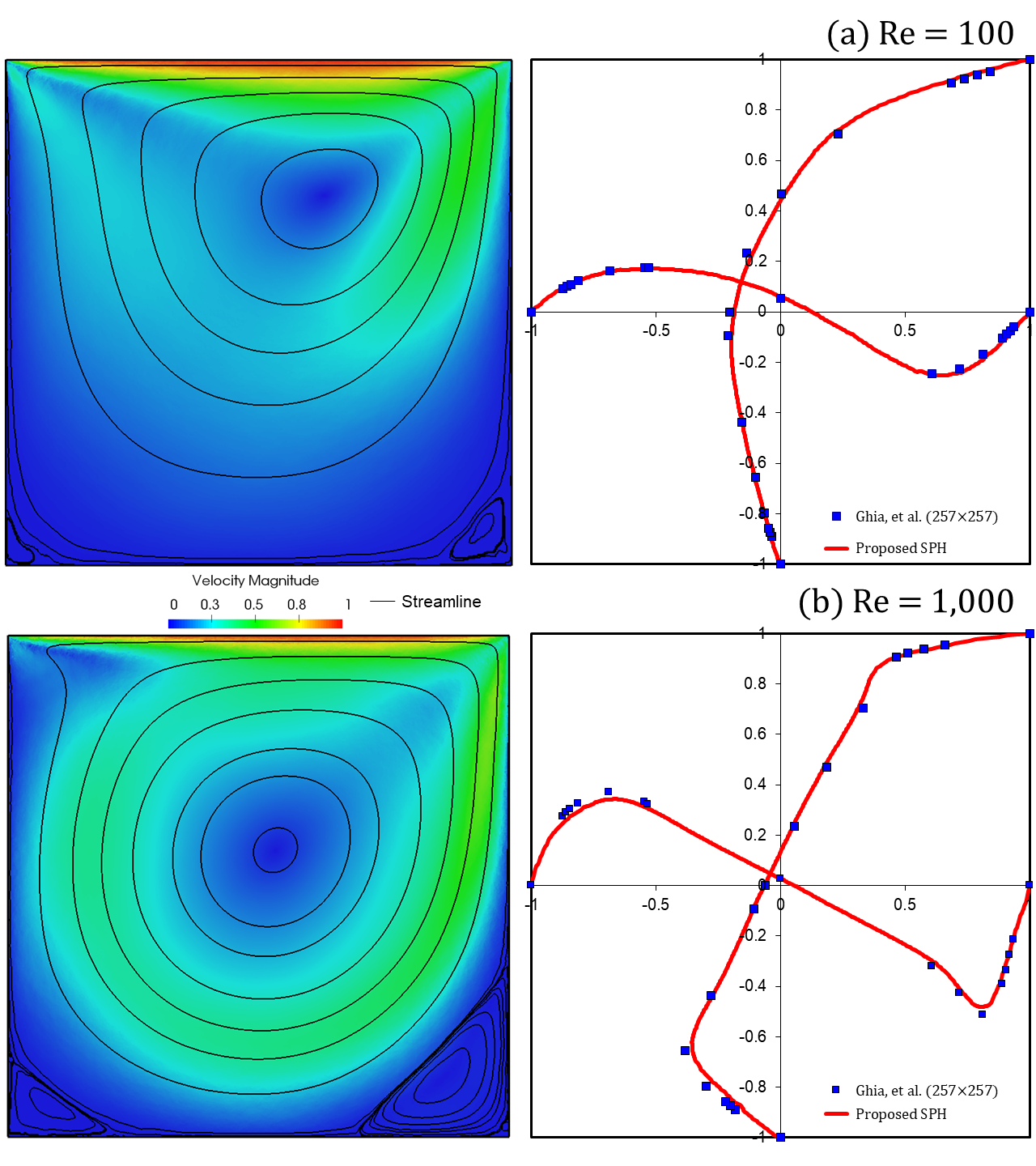}
%\centerline{\includegraphics[scale=0.25]{Figure_ThetaAmp.eps}}
%\end{center}
\caption{Comparisons of driven flow simulations in a square cavity in the cases of (a) $\rm Re=100$ and (b) $\rm Re=1,000$, with the benchmark solutions by Ghia et al.}
\label{fig:CavityFlow}
\end{figure}

\HLLLLL{
To validate the accuracy of our SPH model in respect to normal fluid motions, we simulated the driven flow in a square cavity of size $\rm 1.0~m \times 1.0~m$ by setting the velocity to $\rm 1.0~m/s$ at the top in the horizontal direction and $\rm 0$ otherwise on the boundaries. Here, we introduce the Reynolds number defined as $Re := v_{c}L_{c}/\nu_{c}$, where $v_{c}$, $L_{c}$, and $\nu_{c}$ represent the characteristic values of the speed, length, and kinetic viscosity, respectively. We simulated the two cases of $Re$~=~$\rm{100}$ and $Re$~=~$\rm{1,000}$, assuming the cases closest to the numerical experiments to be discussed in Section~\ref{seq:rotcylinder}. Figure~\ref{fig:CavityFlow} compares the results of the simulations when (a) $Re$~=~$\rm{100}$ and (b) $Re$~=~$\rm{1,000}$ with the benchmark solutions by Ghia et al.~\cite{GHIA1982387} We obtained good agreement in the velocity profiles for both cases with different Reynolds numbers, as shown in Fig.~\ref{fig:CavityFlow} (a) and (b) on the left. Additionally, the streamlines were confirmed to show the characteristics of the driven flow problem, as shown in Fig.~\ref{fig:CavityFlow} (a) and (b) on the right; the small secondary vortices at the bottom of the simulation domain emerge when $Re$~=~$\rm{100}$ and become larger when $Re$~=~$\rm{1,000}$, similar to the benchmark solutions~\cite{GHIA1982387}.

The reason we examined the two cases of $Re$~=~$\rm 100$ and $Re$~=~$\rm~1,000$ is as follows. In Section~\ref{seq:rotcylinder}, we present the numerical experiments we carried out by rotating $\rm ^{4}He$; this was done by setting the outer diameter of the circular vessel to $\rm 0.2~cm$, the temperature $T$ to $\rm 1.6~K$, and by setting the angular velocity of rotating around the axis of the cylinder to $\rm 5~rad/s$. It is difficult to estimate the Reynolds number of the total fluid system because only the normal fluid components have viscosity. Hence, we considered ideal cases in which the simulation domain is only filled with normal fluid. In this case, the Reynolds number is approximately $\rm 176.6$ when setting the abovementioned diameter, temperature, angular velocity, and kinetic viscosity $\nu_{n}$ (=$\mu_{n}/\rho_{n}$) at $T~\rm{=1.6~K}$ to the parameters $v_{c}$, $L_{c}$, and $\nu_{c}$. We also estimated $Re$ to be approximately $\rm 1064.9$ when evaluating $\nu_{c}$ by the parameter $\nu$, which is equal to $\mu_{n}/\rho$. We then take up the closest cases of $Re$~=~$\rm~100$ and $Re$~=~$\rm 1,000$ to these estimations, among those that can be compared with the benchmark solutions.
}

\HL{\section{Numerical Analyses} \label{seq:numericans}}
In this section, we demonstrate our method by performing three major numerical tests: the Rayleigh--Taylor instability, entropy wave propagation analyses, and rotating cylinder simulations.

In Section~\ref{seq:RTI}, \HLLLLL{we carry out a Rayleigh--Taylor instability analysis to obtain a fluid-mechanical description of the interactions between two components when simultaneously solving both equations of motion of the two-fluid model.} \HLLLLL{Recently,} the linear growth of the mushroom structure similar to that in classical fluids has been reported to be observed in the direct simulation of the Gross-Pitaevskii equation \HLLLLL{in the case of two-component BEC}~\cite{PhysRevA.80.063611}. \HLLLLL{Because of the similarity of the systems, we expect to observe a similar type of instability in our case}. We examine whether we can find the emergence of the linear growth of the mushroom structure on the condition of \HLLLLL{a} separate state of superfluid and normal fluid.
\HL{
Meanwhile, simulating wave propagation is challenging for the SPH, particularly in the case of an explicit time-integrating scheme. We discuss the applicability of SPH to entropy wave propagation problems by examining the effect of density fluctuations on the error in calculations with a hypothetical case in Section \ref{seq:wea}.	
}

Furthermore, we perform rotating cylinder simulations, as discussed in Section~\ref{seq:rotcylinder}. In this test, the superfluid is rotated by a cylindrical vessel in a single direction at a constant angular velocity. A similar setup has been reported in previous experimental studies~\cite{PhysRev.60.356, Yarmchuk1982}; these studies noted interesting phenomena that were different from those in classical fluid cases. Specifically, multiple quantum vortices parallel to the cylinder axis emerged and rotated around their respective axes in the same direction as the cylindrical vessel. These vortices were spontaneously aligned, forming a lattice, the so-called ``quantum lattice,'' and revolving around the cylindrical axis while maintaining constant relative positions with each other, similar to rigid body rotation. 
\HLLLLL{Here we investigate the possibility of reproducing several quantum mechanical phenomena observed in liquid $\rm ^{4} He$ even when solving the phenomenological governing equations of liquid $\rm ^{4} He$ using our SPH method.}
%Our present model does not consider any quantum effects. Therefore, it might not be easy to reproduce these phenomena. Nevertheless, we perform rotating cylinder simulations with the aim of clarifying discrepancies between simulations and identifying the improvement in our model.

\begin{figure*}[t]
% \begin{figure}[t]
\vspace{-10.0cm}
%\vspace{-0.3cm}
\begin{center}
\includegraphics[width=2.1\textwidth, clip, bb= 0 0 2031 1127]{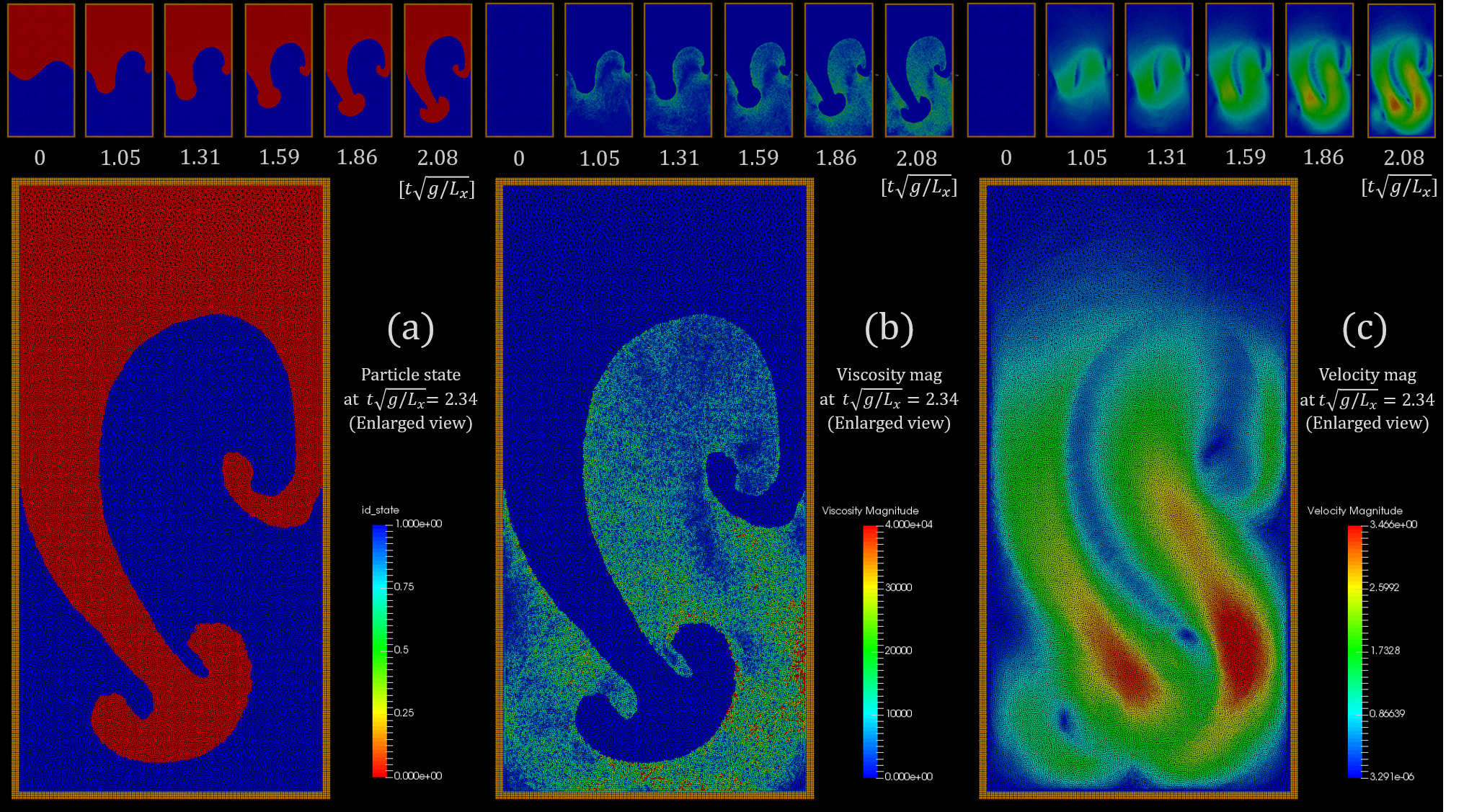}
%\includegraphics[width=1.0\textwidth, clip, bb= 0 0 2031 1127]{Figure_RTISnapshot.png}
% \centerline{\includegraphics[scale=0.5]{Figure_RTISnapshot.eps}}
%\centerline{\includegraphics[scale=0.5]{Figure_RTISnapshot.pdf}}
\end{center}
\caption{\HL{Snapshots of the Rayleigh--Taylor instability simulation until the emergence of linear growth of the mushroom structure when setting $(N_{px}, N_{py})$ to be $\rm (240, 480)$ and setting the density ratio to be $\rho_{s}/\rho_{n} = 5.024$.}}
\label{fig:RTI:snapshot}
%\end{figure}
\end{figure*}

\HLLLLL{\subsection{Rayleigh--Taylor instability analysis}\label{seq:RTI}}
\HL{
We set the simulation domain of $(L_{x}, L_{y})$ to be $\rm (100~{\mu m}, 200~{\mu m})$ and place the superfluid particles at $y > 0.5L_{y}$ and the normal fluid particles at $y < 0.5L_{y}$. We set the resolution of $(N_{px}, N_{py})$ to be $\rm (240, 480)$, where $N_{px}$ and $N_{py}$ are the number of particles arranged in each direction. The interface between the two phases is disturbed by the function $y = L_{x}(1.0-0.15~{\rm sin}(2.0\pi x/L_{x}))$, and the value of the gravity constant of $\rm 9.8~m/s^{2}$ is set in the y-direction, in a manner similar to the case in a classical fluid~\cite{Shadloo2013}. We arrange multiple layers of frozen particles as walls, alongside the boundary of the simulation domain. No-slip conditions are imposed on the wall particles as boundary conditions.}~This is acceptable because both normal and superfluid components have frictional forces against the normal fluid particles forming the wall.

\HL{The reference temperature $T_{0}$ is set to be $\rm 1.6~K$. 
% Using $T_{0}$. we set the temperature of superfluid particles to be $T_{0} \times 0.99$ and set that of normal fluid particles to be $T_{0} \times 1.01$. 
The parameters ($\rho_{s}$, $\rho_{n}$, $\alpha$, $\beta_{v}$, $\kappa$, $\eta_{n}$) are determined by reference to the values at $T_{0} = {\rm 1.6~K}$ in Table~$1$ and Table~$2$ in the literature~\cite{Vinen2002}. The resulting density ratio of $\rho_{s}/\rho_{n}$ becomes $5.024$. 
Meanwhile, we set the vortex line density $L_{0}$ to be $\rm 1 \times 10^6~cm^{-1}$ as an initial state, in reference to the literature~\cite{NEMIROVSKII201385}. 
The parameters ($\bar{\eta}$, $\bar{\eta}_{r}$, $C_{\omega}$) are set to be ($\eta_{n}$, 0, 0) in this test. 
We estimate the dimensionless parameter $C_{e}$ to be $1.428\times 10^{-3}$ considering the volume ratio of $V_{\rm liquid}/V_{\rm air}$ at $\rm 1.01325~bar$~\cite{hammond2000elements}. The value of $NM$ in Eq.~(\ref{eq:entrodens}) is set equivalent to $\rho_{n} {L_{x}}^{2}$ because only the normal fluid components carry entropy.}

\HL{Figure~\ref{fig:RTI:snapshot} shows snapshots of the Rayleigh--Taylor instability simulation. The red part and blue part in Fig.~\ref{fig:RTI:snapshot}(a) respectively represent the superfluid particles and normal particles. As a result, we confirmed the emergence of the characteristic mushroom structure in the linear growth regime. The mushroom structure shows its growth in the slanting direction because of wall boundary conditions; this is the same as the case for a classical fluid~\cite{DeGruyter2012}. The time from the start to the final snapshot is approximately $7.49~{\rm ms}$ in physical time, which corresponds to approximately $2.34$ in the time scaled by the factor $\sqrt(g/L_{x})$. Figure~\ref{fig:RTI:snapshot}(b) represents the magnitude of the viscosity term calculated using Eq.~(\ref{eq:laplaciandisc}); the magnitude of viscosity is confirmed to be zero in normal fluid components. Fig.~\ref{fig:RTI:snapshot}(c) shows the magnitude of the velocity of particles for reference.}
\HLLLLL{
Consequently, we obtained the description of the interactions between two components when simultaneously solving both equations of motion of the two-fluid model; it was observed to show the linear growth of the mushroom structure unique to the Rayleigh--Taylor instability problems, similar to ordinary fluids.}\\

\begin{figure*}[t]
%\vspace{+0.25cm}
\vspace{-0.7cm}
%\begin{center}
\includegraphics[width=1.33\textwidth, clip, bb= 0 0 1123 189]{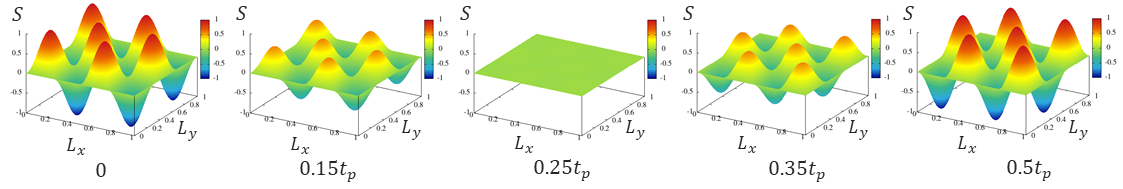}
% \centerline{\includegraphics[scale=0.595]{Figure_WEProblems_Profiles.eps}}
%\end{center}
\caption{\HL{Snapshots of the analytical solution of the targeted problem during the time from $\rm 0~s$ to the half-period of time.}}
\label{fig:wep:profiles}
\end{figure*}
\HL{\subsection{Effect of density fluctuations on the numerical analysis of the entropy wave equation} \label{seq:wea}}
\HL{Let us examine the effect of density fluctuations on the calculation error of wave propagation problems with a hypothetical case.}
\HL{The wave equation is given as}
\HL{
\begin{eqnarray}
\frac{\partial^{2} S}{\partial t^{2}} &=& {c}^{2} \nabla^2 S, \label{eq:waveeq}
\end{eqnarray}
where $c$ is the speed of sound. We impose closed boundary conditions:}
\HL{
\begin{eqnarray}
S(x,0) &=& S(x, L_{y}) = 0, \\
S(0,y) &=& S(L_{x}, y) = 0.
\end{eqnarray}
In this case, the analytical solution of Eq.~(\ref{eq:waveeq}) is given as follows~\cite{kirsch2011math}:}
\HL{
\begin{eqnarray}
S(x,y) &=& A(t)~{\rm sin} \frac{m \pi x}{L_{x}}  {\rm sin} \frac{n \pi y}{L_{y}}
~\biggl( \begin{array}{l}
{m = 1,2,\cdots} \\
{n = 1,2,\cdots} \\
	  \end{array} \biggl),
	\label{eq:we:ans}
\end{eqnarray}
where the time-dependent function $A(t)$ is expressed as}
\HL{
\begin{eqnarray}
A(t) &=& C_{m, n} {\rm sin}\lambda_{m,n} t + C_{m, n}^{\ast} {\rm cos}\lambda_{m,n} t, \\
	   \lambda_{m,n} &=& c\pi\sqrt{\frac{m^2}{L_{x}^2} + \frac{n^2}{L_{y}^2}}.
\end{eqnarray}
Here, $C_{m, n}$ and $C_{m, n}^{\ast}$ correspond to the coefficients of the sine terms and cosine terms in the double Fourier expansion of Eq.~(\ref{eq:we:ans}), respectively. 
In this benchmark test, we discretize the left-hand side of Eq.(\ref{eq:waveeq}) using the second-order central difference method~\cite{iserles_2008} while replacing the right-hand side with the Laplacian operators of SPH or the finite-difference method~(FDM); we can focus on the problem of numerical error in the spatial direction. }
\begin{figure}[t]
\vspace{-1.7cm}
%\begin{center}
\includegraphics[width=0.65\textwidth, clip, bb= 0 0 960 720]{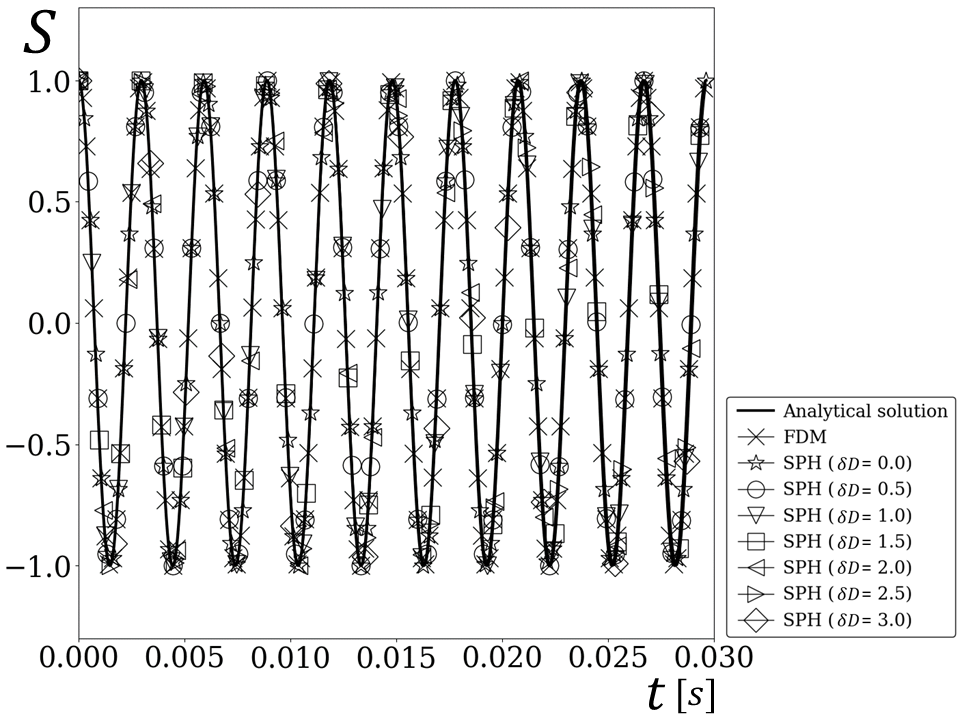}
%\centerline{\includegraphics[scale=0.32]{Figure_OscillationCs135.eps}}
%\end{center}
\caption{\HL{Variation of $S$ at a certain point during the simulation with $C_{s} =135$ in the cases of FDM or SPH with different values of $\delta D$.}
\label{fig:wep:oscillationCs135}}
\end{figure}

\HL{We set the simulation domain of $(L_{x}, L_{y})$ to be $\rm (1.0, 1.0)$ and set the resolution of $(N_{px}, N_{py})$ to be $\rm (400, 400)$.
We take two cases, setting the speed of sound $c$ to be $\rm 135~ms^{-1}$ and $\rm 20~ms^{-1}$, assuming two representative cases of the second sound waves of $\rm ^{4} He$ in the low-temperature region, as mentioned in Section~\ref{sec:calcentro}. We set the pair of $\rm (m, n)$ to be $\rm (4, 3)$ and set the initial velocity of each particle to be $0$ everywhere in the simulation domain. In this case, the values of $C_{m,n}$ and $C_{m,n}^\ast$ become $0$ and $1$.}
\HL{Figure~\ref{fig:wep:profiles} shows snapshots of the analytical solution of the targeted problem during the time from $\rm 0~s$ to the half-period of time.}
\begin{figure}[t]
\vspace{-1.4cm}
%\vspace{+0.3cm}
%\begin{center}
\includegraphics[width=0.65\textwidth, clip, bb= 0 0 1280 720]{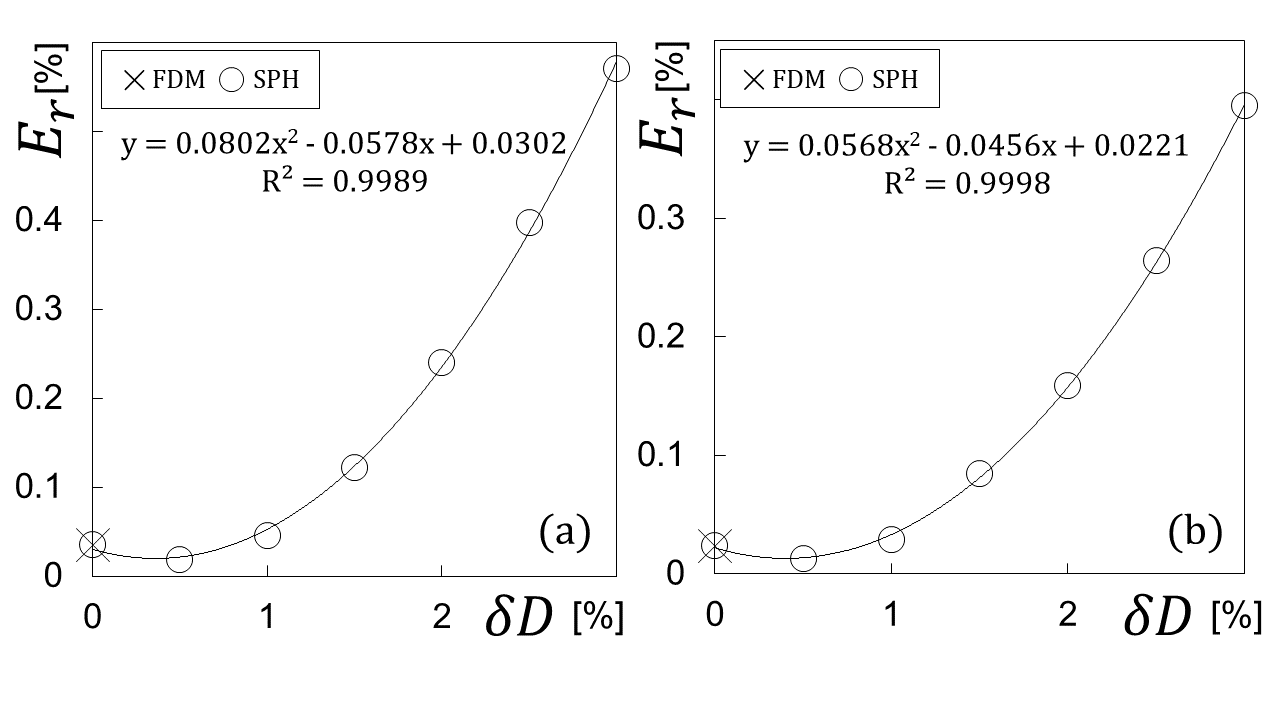}
%\centerline{\includegraphics[scale=0.25]{Figure_DensityFluctuation.eps}}
%\end{center}
\caption{\HL{Dependence of the relative error $E_{r}$ on the parameter $\delta D$: (a) for $C_{s}=135$ and (b) $C_{s} =20$.}}
\label{fig:wep:flucCs135}
\end{figure}

\HLLLLL{In this test}, we set the same mass for all the particles. In this case, only the non-uniformity of particle arrangements contributes to the fluid density fluctuation because of the weighted calculation according to the positions of particles. Let us introduce the initial distance between two particles, $l$, and a distance parameter $dl$. Consider the case where we isotropically arrange particles with an interval of $l$ in each direction and shift the particles by $dl$ in an arbitrary direction. In this case, the averaged small volume $\Delta V_{0}$ is $l^{2}$, and the maximum one $\Delta V$ is $(l+dl)^{2}$. We define a volume fluctuation rate $\delta D$ as $((\Delta V)-\Delta V_{0})/\Delta V_{0} \times 100$. Because of the uniformity of mass and the relationship $\Delta V = m/\rho$, it is possible for us to regard $\delta D$ as a criterion for density fluctuation. 
\HL{Additionally, the physical time is set to $\rm 0.03~s$, roughly assuming the time from the emergence of the second wave to the decay of the vortex line density~\cite{NEMIROVSKII201385}. 
In addition, we introduce a stabilization technique presented in the MPS (moving particle semi-implicit)~\cite{koshizuka1996moving} to suppress the increase in variance caused by repeated calculations (see Appendix~\ref{apend:reformMPS}). We then compare the simulation results of SPH with that of FDM and the analytical solutions given by Eq.~(\ref{eq:we:ans}).}
% $((l+dl)^{2}-l^{2})/l^{2}\times 100$, which is also rewritten as
%\begin{figure}[b]
%\vspace{-0.3cm}
%\begin{center}
%\centerline{\includegraphics[scale=0.3]{Figure_OscillationCs20.eps}}
%\end{center}
%\caption{test.}
%\label{fig:wep:oscillationCs20}
%\end{figure}

\HL{Figure~\ref{fig:wep:oscillationCs135} shows the variation of $S$ at a certain point during the simulation with $C_{s} =135$, in the respective cases. The solid line indicates the analytical solution, and the solid line with the cross symbol shows the simulation result of FDM. The remaining seven lines with symbols indicate the simulation results of SPH with different values of $\delta D$ between $0$ and $3$ per cent. All results were confirmed to show excellent agreement at this viewing scale. }

\HL{Fig.~\ref{fig:wep:flucCs135}(a) shows further analyses of the results in Fig.~\ref{fig:wep:oscillationCs135}, demonstrating the dependence of the relative error on the parameter $\delta D$ in the respective cases. Here, the relative error $E_{r}$ is defined as the average of $|S_{i} - S_{a}|/|S_{a} + A|\times 100$ over all the particles, where $S_{i}$ is the value of $S$ of the $i$th particle and $S_{a}$ is the analytical value of the point on the particle. $A$ is the amplitude of the oscillation, which is equivalent to $1$ in this case and is added to the denominator to avoid division by zero.}
\HL{The solid curve indicates the fitting curves using the least squares method.
The relative error $\delta D$ in the case of SPH was found to be a minimum at approximately $\delta D = 0.5$; subsequently, it increased as $\delta D$ increased by a power of two. 
Meanwhile, Fig.~\ref{fig:wep:flucCs135}(b) shows the simulation results in the case where $C_{s} = 20$. The dependence of $E_{r}$ on the parameter $\delta D$ shows the same increase as in the case where $C_{s} = 135$.}

\begin{figure}[t]
\vspace{-1.8cm}
%\vspace{+0.2cm}
%\begin{center}
\includegraphics[width=0.6\textwidth, clip, bb= 0 0 960 720]{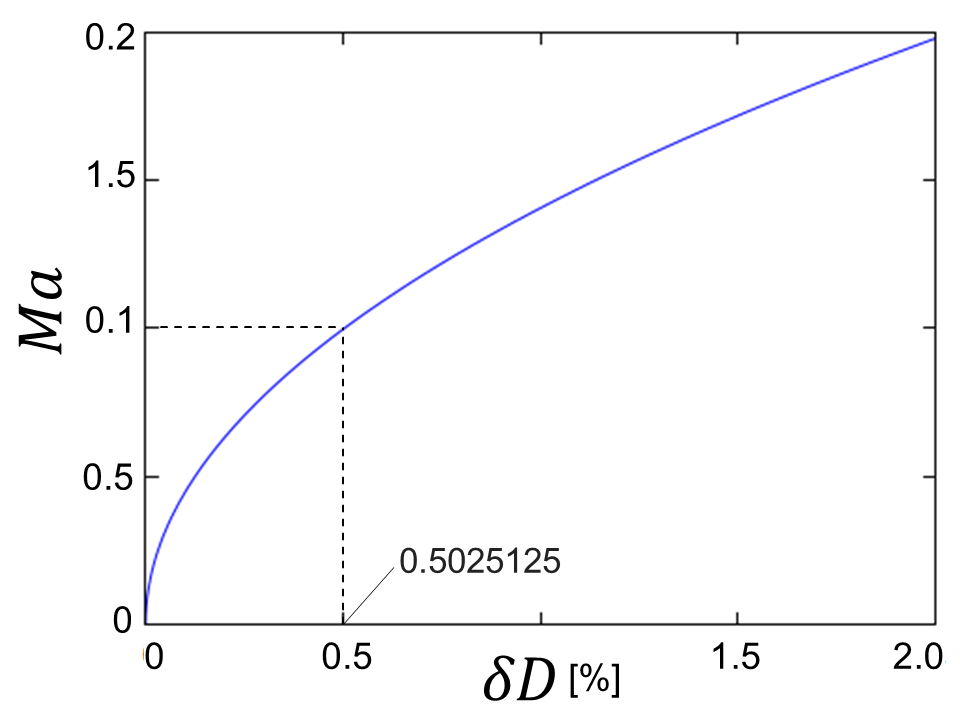}
%\centerline{\includegraphics[scale=0.3]{Figure_RelationMandD.eps}}
%\end{center}
\caption{\HL{Relationship between the mach number $Ma$ and the parameter $\delta D$.}}
\label{fig:wep:relationMD}
\end{figure}
\begin{figure*}[t]
\vspace{-6.3cm}
%\vspace{-0.3cm}
\begin{center}
\includegraphics[width=2.1\textwidth, clip, bb= 0 0 2014 666]{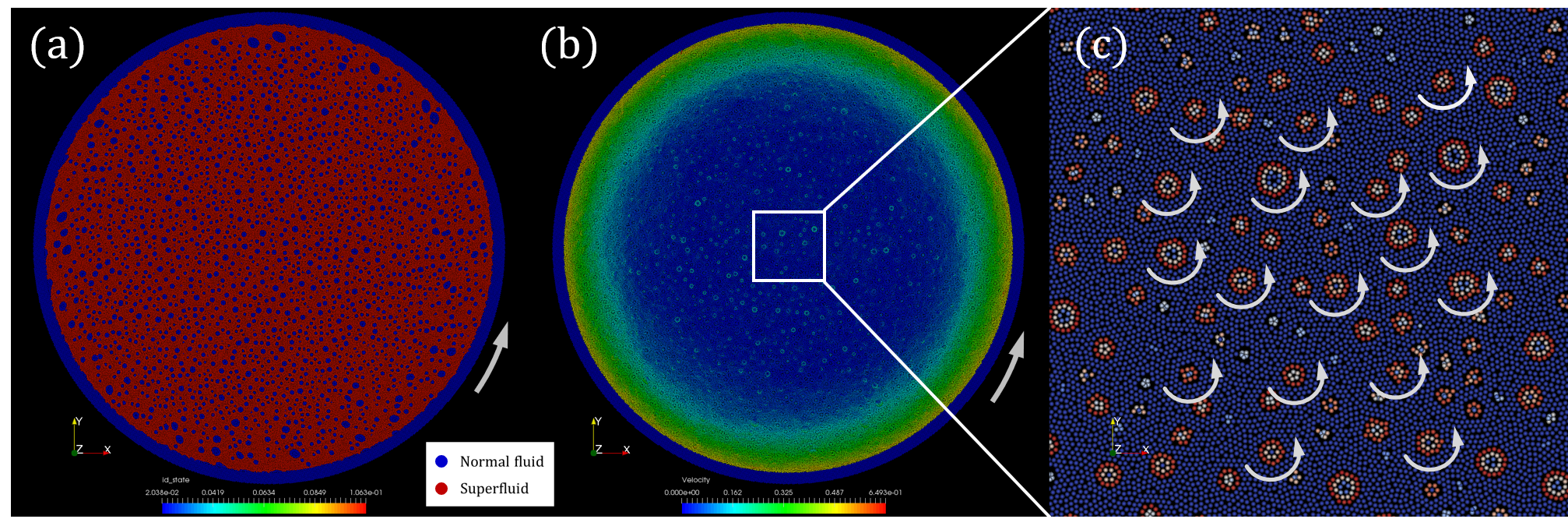}
\end{center}
\caption{An image of the rotating cylinder simulation at $t {\rm = 0.5~s}$: (a) a view representing the superfluid particles in red and normal fluid particles in blue; (b) a view colored by the \HLLLLL{magnitude of} velocit\HLLLLL{ies} of the particles; and (c) an enlarged view of the center of (b) colored by the magnitude of rotational forces given by Eq.~(\ref{eq:rotforce}). A movie of the complete simulation is provided as supplemental material, and its online version is available at https://www.satoritsuzuki.org/video-preprintarxiv-200411052.}
\label{fig:RotCylinSim:snapshot}
\end{figure*}

\HL{The results in Fig.~\ref{fig:wep:flucCs135} demonstrate the applicability of SPH to wave propagation problems of liquid $\rm ^{4} He$, even though the relative error $E_{r}$ depends on the parameter $\delta D$ by a power of two. This occurs because the relative error is small enough to be less than $0.25$ per cent, even in the case where $\delta D = 2$. Here, $\delta D$ is known to be approximately $0.5$, in usual cases. Thus, the relative error is approximately $0.012$ per cent for $C_{s} = 20$, and $0.018$ per cent for $C_{s} = 135$. Through the discussion in this section, it is shown that we can perform wave propagation analyses within the range of these permissible errors.}

\HL{The reason why $\delta D$ is estimated to be $0.5$ is explained as follows. The Mach number, the ratio of the maximum velocity to the speed of sound, is denoted as $M_{a}$. Given that the temperature is constant, the relative error of density is proportional to the square of $M_{a}$, which is described as $(\rho_{0}-\rho)/\rho_{0} = 0.5M_{a}^{2}$ as in~\cite{MORRIS199741}. Because of the uniformity of mass, we can obtain the relationship $2(1-\Delta V_{0}/\Delta V) = M_{a}^{2}$, where $\Delta V_{0}=m/\rho_0$ and $\Delta V = m/\rho$. Because $\Delta V_{0}$ is $l^{2}$ and $\Delta V$ is $(l+dl)^{2}$, a simple calculation leads to the following relationship between $\delta D$ and $M_{a}$:}
\begin{eqnarray}
M_{a} = \sqrt{2(1.0-\frac{1}{\delta D/100 + 1.0})}. \label{eq:relMandD}
\end{eqnarray}
Figure~\ref{fig:wep:relationMD} shows the relationship between $M_{a}$ and $\delta D$ described by Eq.~(\ref{eq:relMandD}). 
In the simulation of incompressible SPH, $M_{a}$ is given as an input parameter, which is typically set to $0.1$. The value of $\delta D$ at this time is approximately $0.5$, as shown in Fig.~\ref{fig:wep:relationMD}.\\

\subsection{Rotating cylinder simulations}\label{seq:rotcylinder}
The outer diameter of the circular vessel was set to be $\rm 0.2~cm$. We set the angular velocity of rotating around the cylinder axis to $\rm 5~rad/s$ in the counterclockwise direction. The resolution of ($N_{px}$, $N_{py}$) was set to be $\rm (500, 500)$. We set the temperature $T_{0}$ to $\rm 1.6~K$ and determined the parameters ($\rho_{s}$, $\rho_{n}$, $\alpha$, $\beta_{v}$, $\kappa$, $\eta_{n}$) by referring to the values at $T_{0} {\rm = 1.6~K}$ listed in a previous report~\cite{Vinen2002}. Meanwhile, the parameters ($\bar{\eta}$, $\bar{\eta}_{r}$, $C_{\omega}$) are set to be ($\eta_{n}$, $\eta_{n}$, 0.01) in this simulation test. The wall particles, made of normal fluid particles, are arranged along the inner circumference of the vessel. No-slip conditions are imposed on the wall particles as boundary conditions. This is acceptable because both normal and superfluid components have frictional forces against the normal fluid particles forming the wall. In the initial state, superfluid or normal fluid particles are randomly generated in proportion to the density ratio and are distributed inside the vessel.

Figure~\ref{fig:RotCylinSim:snapshot} shows an image of the rotating cylinder simulation at $t~{\rm = 0.5~s}$. In this figure, (a) represents the superfluid particles in red and normal fluid particles in blue; (b) shows a view colored by the \HLLLLL{magnitude} of velocit\HLLLLL{ies} of the particles; and (c) shows an enlarged view of the center of (b) colored by the magnitude of rotational forces given by Eq.~(\ref{eq:rotforce}). A movie of the complete simulation is provided as supplemental material, and its online version is available at the link \HLLLLL{in the caption of Fig.~\ref{fig:RotCylinSim:snapshot}}. This movie has a viewing time of 36~s, which corresponds to 5~s in terms of physical time.

Interesting observations were noted. Moments after commencing the simulation, normal fluid particles gathered and formed clusters, which changed to vortices revolving around their individual axes, with their pores filled with normal components. These vortices changed their locations by interacting with each other, while occasionally connecting with smaller vortices or absorbing the normal particles that failed to form vortices. The vortices located near walls are affected by the forced rotation of the outer vessel. Because the effective range of forced rotation continuously increases in space as physical time increases, the vortices gradually rotate around the circular axis, independent from their own rotations.
\begin{figure}[t]
\vspace{-1.8cm}
%\vspace{+0.3cm}
%\begin{center}
\includegraphics[width=0.65\textwidth, clip, bb= 0 0 1280 720]{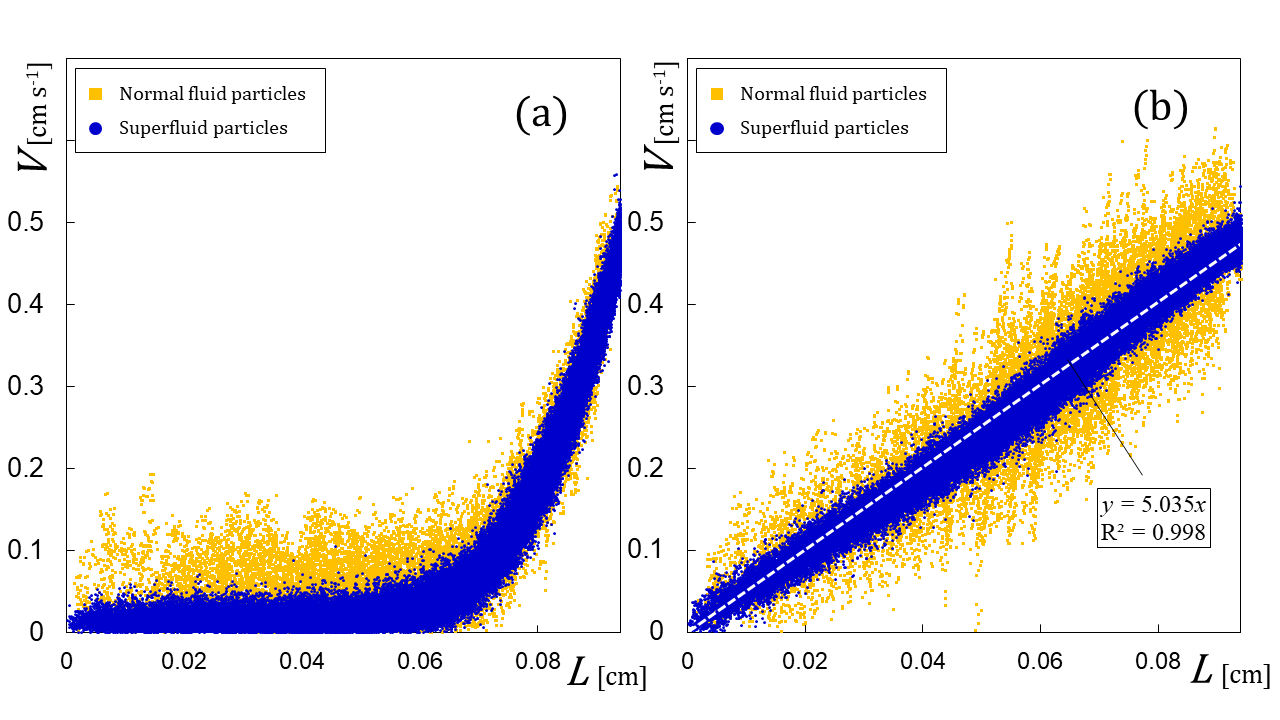}
%\centerline{\includegraphics[scale=0.25]{Figure_DensityFluctuation.eps}}
%\end{center}
\caption{Distributions of the velocity magnitude of all particles at (a) $t = {\rm 0.5~s}$ and at (b) $t = {\rm 5~s}$. }
\label{fig:RotCylinSim:veldist}
\end{figure}

Figure~\ref{fig:RotCylinSim:veldist} shows the distributions of the velocity of all particles at (a) $t = {\rm 0.5~s}$ and (b) $t = {\rm 5~s}$. The horizontal line indicates the distance between a particle and the center of the circular axis, and the vertical line represents particle velocity. From Fig.~\ref{fig:RotCylinSim:veldist}(a), it is confirmed that the velocities of particles located in the vicinity of the wall are considerably greater than those of particles in the inner area because of the effect of the forced rotation. The velocities of normal fluid particles in the inner area at $L < {\rm 0.06~cm}$ are confirmed to fluctuate more widely than those of the superfluid particles, because the normal particles form the vortices and rotate around the axes of the vortices.  As time passes, the vessel's forced rotation gradually affects the inner part of the vessel. Fig.~\ref{fig:RotCylinSim:veldist}(b) shows the velocity distribution for $t {\rm =5~s}$. The white dotted line represents the fitting line obtained by the least squares method. It is understood from (b) that all the vortices revolve around the circular vessel'axis with almost the same angular velocity of $|\vec{\omega}| = {\rm 5~rad/s}$ and that their averaged velocities follow a relation of $L|\vec{\omega}|$. This result suggests that the vortices exhibit rigid-body rotation. 

As mentioned above, previous experiments have reported that multiple quantum vortices parallel to the cylinder axis emerged and rotated around their respective axes in the same direction as the cylindrical vessel. These vortices are arranged such that they form a lattice, called the ``quantum lattice.'' They revolve around the circular axis while maintaining constant relative positions with each other, similar to rigid body rotation. Through comparisons of our simulation results and the experimental results, it can be said that our current model succeeds in reproducing the first and third phenomena, i.e., the emergence of multiple independent vortices parallel to the circular axis and that of ``rigid-body rotation.''

\begin{figure}[t]
\vspace{-8.5cm}
%\vspace{+0.3cm}
%\begin{center}
\includegraphics[width=0.9\textwidth, clip, bb= 0 0 1125 1125]{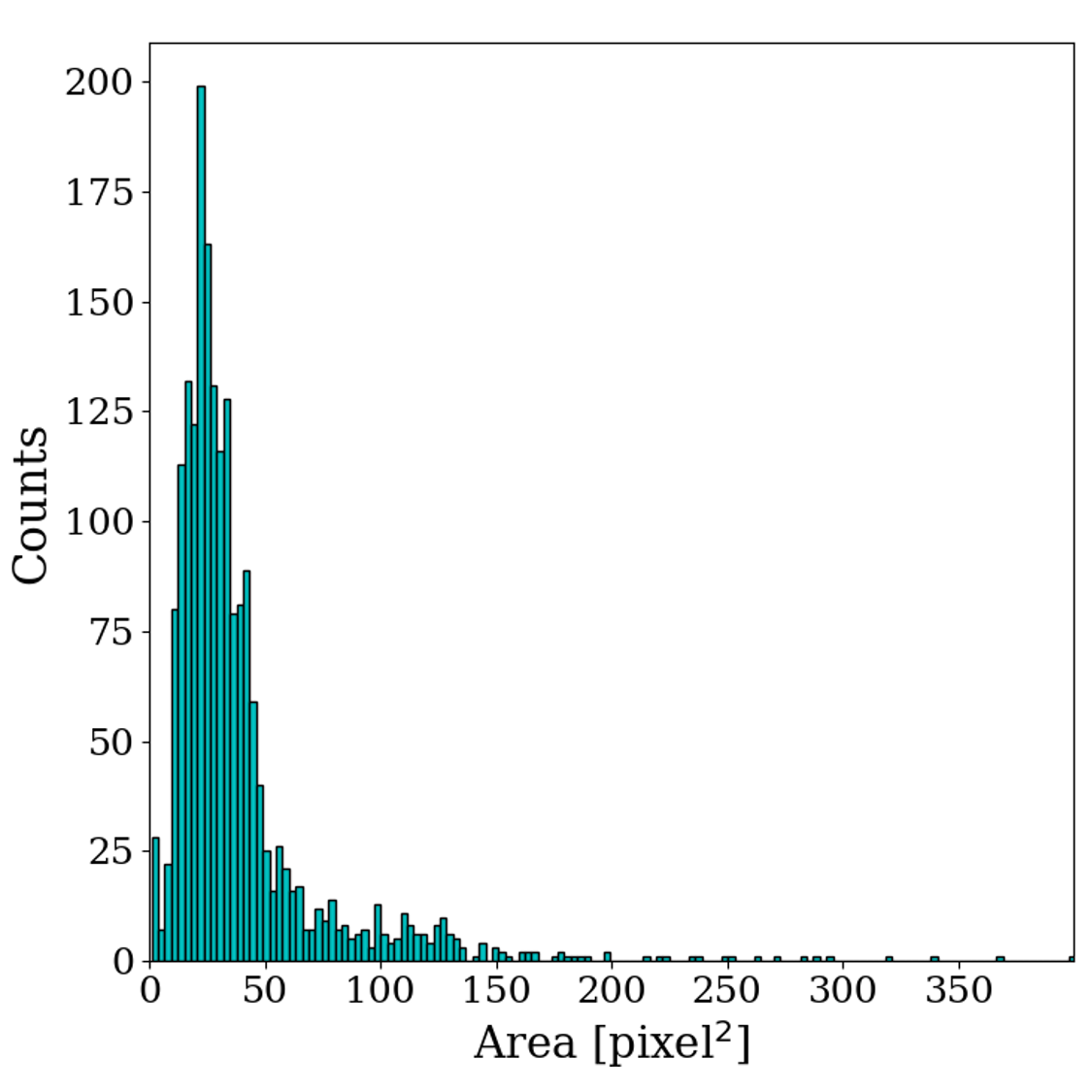}
%\centerline{\includegraphics[scale=0.25]{Figure_DensityFluctuation.eps}}
%\end{center}
\caption{Histogram of the pore areas in the vortices for $t = {\rm 5~s}$.}
\label{fig:RotCylinSim:AreaHist}
\end{figure}

Unfortunately, our current model needs improvements in terms of reproducing quantum lattices. It would be necessary to generate uniform vortices to ensure that the repulsive forces among them balance each other, thereby forming a lattice. Figure~\ref{fig:RotCylinSim:AreaHist} shows a histogram of the pore areas in the vortices for $t = {\rm 5~s}$, which is obtained from the resulting image using {\tt ImageJ}~\cite{10030139275, abramoff2004image, Schneider2012}. $\rm pixel^{2}$ gives the unit of area. The histogram visually clarifies the discrepancy between the target phenomena and our simulations; the histogram should tend to a delta function as the vortices become uniform. A promising strategy to realize such convergence is to incorporate the ``quantization of circulation'' in SPH formalism to ensure that each vortex has the same size and energy; this needs to be studied in future works.

Additionally, we performed the same simulation test for different values of $C_{\omega}$. It was confirmed that the magnitude of $C_{\omega}$ determines the balance between the strongness of vortices and that of forced rotation. At approximately $C_\omega > {\rm 0.015}$, the vortices swell by connecting with or absorbing each other. Consequently, several enlarged vortices continue moving inside the simulation area, almost regardless of the vessel's forced rotation. In contrast, the vortices hardly rotate around  their own axes at approximately $C_\omega < {\rm 0.01}$.

It has been assumed that only a single vortex emerges when addressing similar problems without considering quantum mechanics. However, our simulation results show that the two major phenomena of rotating liquid $\rm ^{4} He$, i.e., the emergence of multiple independent vortices parallel to the circular axis and that of the so-called ``rigid-body rotation,'' can be reproduced by solving the two-fluid model, which is the phenomenological governing equation of liquid $\rm ^{4}He$, by using SPH. The rotational cylinder problem is one of the few problems exhibiting different phenomena depending on basic mechanics or quantum mechanics. It is meaningful that our model succeeded in reproducing the phenomena observed in quantum cases, even though we solve the phenomenological governing equations of liquid $\rm ^{4}He$.

\section{Conclusions} \label{sec:conc}
In this study, we theoretically derived a finite particle approximation of the two-fluid model of superfluid $\rm ^{4}He$ using smoothed particle hydrodynamics. 
To solve the governing equations, we introduced an auxiliary equation representing the microscopic relationship between entropy and temperature, which was derived from quantum statistical mechanics in elementary particle physics. 
\HLLLLL{We then formulated a new SPH model that simultaneously solves both equations of motion of the two-fluid model.}
\HLLLLL{In particular, we presented a reformulation of the viscosity term in the two-fluid model to conserve the angular momentum of the fluid particles around their axes.} Through numerical tests, several sophisticated techniques have been introduced. \HLLLLL{Specifically}, we utilized a Riemann solver among neighboring particles to stabilize the simulation as an alternative to artificial viscosity. We demonstrated a Rayleigh--Taylor instability simulation\HLLLLL{,} confirming the emergence of the characteristic mushroom structure in the linear growth regime. We investigated the applicability of SPH to entropy wave propagation analyses by examining the effect of density fluctuations on the errors in calculations with a benchmark case. The simulation results show good agreement with those of the FDM and the analytical solution.

Furthermore, we performed a simulation of rotating liquid $\rm ^{4}He$. It was found that two major phenomena of rotating liquid $\rm ^{4}He$---the emergence of multiple independent vortices parallel to the circular axis and that of the so-called ``rigid-body rotation''---can be reproduced by solving the two-fluid model, which is the phenomenological governing equation, by using SPH. This finding is interesting because it was previously assumed that only a single vortex emerges when addressing similar problems without considering quantum mechanics. 
\HLLLLL{Our further analysis found that the emergence of multiple independent vortices can be realized by the aforementioned reformulation of the viscosity term to conserve the angular momentum of the particles around their axes.}
It is meaningful that our model succeeded in reproducing the phenomena observed in quantum cases, even though we solve the phenomenological governing equations of liquid $\rm ^{4}He$.

Consequently, our simulation model has been confirmed to reproduce the phenomenological characteristics of superfluid $\rm ^{4}He$. We have, thus, pioneered a new paradigm of Lagrangian particle mechanics, which will hopefully stimulate further numerical studies on superfluids in \HLLLLL{low-temperature physics}. 

\HLLLLL{As mentioned in Section~\ref{seq:rotcylinder}}, a promising strategy to reproduce \HLLLLL{quantum lattices} is to incorporate the ``quantization of circulation'' in SPH formalism to ensure that each vortex has the same size and energy; this needs to be studied in future works. \HLLLLL{Furthermore}, it is \HLLLLL{more accurate} to introduce the model of state transition between superfluid and normal fluid particles since it is induced by several types of instability unique to superfluid $\rm ^{4}He$. In such cases, each particle could stochastically change its state according to quantum statistical mechanics; the most straightforward approach is to estimate the probability of transition using the distribution of Bose-Einstein condensates. The suitable transition models should also be discussed in future studies.

\section*{Acknowledgement}
The authors thank H.~Tsuzuki for his constructive comments.
This research was supported by JSPS KAKENHI Grant Number 19K21528, and partly supported by ``Grant-in-Aid for JSPS Fellows'' in Japan.
The author would like to thank Editage (www.editage.jp) for English language editing.
The author would also like to express his gratitude to his family for their moral support and warm encouragement.

\appendix
\section{A reformulation of the Laplacian operator on the basis of MPS formalism} \label{apend:reformMPS}

\HL{Equation~(\ref{eq:laplacian}) is reposted as}
\HL{
\begin{eqnarray}
	\nabla^{2} \phi(\vec{r}_{i}) 
		= \sum^{N_{p}}_{j} \frac{m_{j}}{\rho_{j}}\frac{\phi(\vec{r}_{i})-\phi(\vec{r}_{j})}{|\vec{r}_{ij}|^{2}}  
	\vec{r}_{ij}\cdot \nabla W_{ij}, \label{eq:laplacian:scalarA}
\end{eqnarray}
where $\vec{r}_{ij} = \vec{r}_{i} - \vec{r}_{j}$. }
% \HL{In the case where all particles have the same mass and density, we can replace $m_{j}/\rho_{j}$ by $m_{i}/\rho_{i}$ in Eq.~(\ref{eq:laplacian:scalarA}) as follows:}
\HL{In the case where all particles have the same mass and density, we can replace $m_{j}/\rho_{j}$ by $m_{i}/\rho_{i}$ in Eq.~(\ref{eq:laplacian:scalarA}) as follows:}
\HL{
\begin{eqnarray}
	\nabla^{2} \phi(\vec{r}_{i}) 
		= \sum^{N_{p}}_{j} \frac{m_{i}}{\rho_{i}}\frac{\phi(\vec{r}_{i})-\phi(\vec{r}_{j})}{|\vec{r}_{ij}|^{2}}  
	\vec{r}_{ij}\cdot \nabla W_{ij}, \label{eq:laplacian:scalarB}
\end{eqnarray}
Consider the functions:}
\HL{
\begin{eqnarray}
{\mathbb P}_{ij} &:=& \frac{w(\vec{r}_{ij})}{\sum _{j} w(\vec{r}_{ij})}, \label{eq:probdef}\\
w(\vec{r}_{ij}) &:=& - \frac{1}{2D_{s}}\vec{r}_{ij}\cdot \nabla W_{ij},
\end{eqnarray}
where $D_{s}$ is the dimension.}
\HL{Equation~(\ref{eq:probdef}) can be regarded as a probability function when $w(\vec{r}_{ij})$ is positive everywhere in the neighborhood of the $i$th particle.
% To give examples, the Gaussian kernel satisfies this condition. 
To give examples, the Gaussian kernel and B-spline kernel satisfy this condition. 
Because of the uniformity of mass, we can select a set of the density and small volume of the $i$th particle that has the relationship as}
\HL{
\begin{eqnarray}
	\rho_{i}&=& \frac{m_{i}}{\Delta V_{i}}, \label{eq:laplacian:scalarC} \\
	\Delta V_{i}&=& \frac{1}{\sum _{j} w(\vec{r}_{ij})}. \label{eq:laplacian:scalarC2}
\end{eqnarray}
From Eq.~(\ref{eq:probdef}) to Eq.~(\ref{eq:laplacian:scalarC2}), Eq.~(\ref{eq:laplacian:scalarB}) is rewritten as}
\HL{
\begin{eqnarray} 
	\nabla^{2} \phi(\vec{r}_{i}) 
		= 2D_{s} \sum^{N_{p}}_{j} \frac{\phi(\vec{r}_{j})-\phi(\vec{r}_{i})}{|\vec{r}_{ij}|^{2}} {\mathbb P}_{ij}. \label{eq:laplacian:scalarD}
\end{eqnarray}
Here, we replace the variance $|\vec{r}_{ij}|^{2}$ by its averaged expectation as}
\HL{
\begin{eqnarray}
	\nabla^{2} \phi(\vec{r}_{i}) 
		&=& \frac{2D_{s}}{\tau} \sum^{N_{p}}_{j} \bigl(\phi(\vec{r}_{j})-\phi(\vec{r}_{i})\bigr) {\mathbb P}_{ij}, \label{eq:laplacian:scalarE} \\
   \tau &=& \sum_{j} |\vec{r}_{ij}|^{2} {\mathbb P}_{ij}. \label{eq:laplacian:scalarE2}
\end{eqnarray}
Note that the final forms of Eq.~(\ref{eq:laplacian:scalarE}) and Eq.~(\ref{eq:laplacian:scalarE2}) are irrelevant to the choices of $\rho_{i}$ and $\Delta V_{i}$.
The introduction of $\tau$ is known to suppress the increase in variance caused by repeated calculations~\cite{koshizuka1996moving, Koshizuka1998}.}
%\\\\\\\\\\\\

\bibliographystyle{h-physrev3}
\bibliography{reference}

\clearpage

%\begin{thebibliography}{00}

%% \bibitem{label}
%% Text of bibliographic item

%\bibitem{}

%\end{thebibliography}
\end{document}